\documentclass{article}

\usepackage{anyfontsize}
\usepackage[utf8]{inputenc}
\usepackage[T1]{fontenc}
\usepackage[sfdefault]{biolinum}
\usepackage[sansmath]{libertinust1math}

\usepackage{amsfonts}
\usepackage{graphicx}
\usepackage{amsmath}
\usepackage{amssymb}
\usepackage{multirow}
\usepackage{braket}
\usepackage{url}
\usepackage{booktabs}
\usepackage{caption}
\usepackage{subcaption}
\usepackage{rotating}
\usepackage[english]{babel}

\usepackage{hyperref}
\usepackage{cleveref}
\hypersetup{
  colorlinks=true,
  linkcolor=blue,
  filecolor=magenta,      
  urlcolor=cyan
}

\newcommand{\llbracket}{{[\![}}
\newcommand{\rrbracket}{{]\!]}}

\newcommand{\algoone}{\cite[Algo.~1]{patel2008optimal}}

\title{Decoding techniques applied to the compilation of CNOT circuits
  for NISQ architectures\footnote{This document is the author's
    version of the corresponding research manuscript prior to formal
    peer review. An updated version is published in Science of Computer
    Programming, Volume 214, 1 February 2022, 102726,
    \href{https://doi.org/10.1016/j.scico.2021.102726}{10.1016/j.scico.2021.102726}.}}

\author{
Timothée Goubault de Brugière$^{1,3}$,
Marc Baboulin$^{1}$, Benoît Valiron$^{2}$,\\
Simon Martiel$^{3}$ and Cyril Allouche$^{3}$}
\date{{\fontsize{10}{15}\selectfont
  $^{1}${\textit{Laboratoire de Recherche en Informatique, Université Paris-Saclay}, 
Orsay, France}\\
$^2${\textit{Laboratoire de Recherche en Informatique, CentraleSupélec}, 
Orsay, France}\\
$^3${\textit{Atos Quantum Lab}, Les Clayes-sous-Bois, France}}}

\begin{document}

\maketitle

\begin{abstract}
  Current proposals for quantum compilers require the synthesis and
  optimization of linear reversible circuits and among them CNOT
  circuits. 
    Since these circuits
    represent a significant part of the cost of running an entire
    quantum circuit, we aim at reducing their size.
  In this paper we present a new algorithm for the synthesis of
  CNOT circuits based on the solution of the syndrome decoding
  problem. Our method addresses the case of ideal hardware
  with an all-to-all qubit connectivity and the case of near-term
  quantum devices with restricted connectivity. For both cases, we present benchmarks showing that
  our algorithm outperforms existing algorithms. %in both cases of partial and full connectivity.
\end{abstract}
 
\section{Introduction}

Quantum compilers transform a quantum algorithm into an optimized
sequence of instructions (elementary gates) directly executable by the
hardware. The most common universal set of gates for this task is the
Clifford+T gate set, used in many quantum architectures
\cite{campbell2017roads}.

For fault-tolerant computation, the T gate is considered to be the
most costly gate to implement fault-tolerantly (using for instance
magic state distillation protocols \cite{campbell2012magic}) and many
efforts have been made to reduce their number in quantum
circuits~\cite{2058-9565-4-1-015004,6899791,kissinger2020reducing}. Yet,
when implementing complex quantum algorithms it is estimated that the
total number of CNOT gates increases much more rapidly with the number
of qubits than the number of T gates, and it is likely that the global
CNOT cost will not be negligible on large sized registers
\cite{2058-9565-4-1-015004,maslov2016optimal} compared to the global T
cost.

For the moment, fault-tolerant quantum computers are not
available. The current devices are medium-sized chips (less than 100
qubits) called NISQ computers (Noisy Intermediate Scale Quantum)
\cite{preskill2018quantum}. Those computers are prone to errors and
especially the two-qubit gates such as the CNOT gate have a much lower
fidelity compared to one-qubit gates, including the T gate (see, e.g.,
the results for Rigetti's Aspen-8 chip \cite{rigetti}). So, for NISQ
processors it is a priority to minimize the number of CNOT
gates. Moreover, for some technologies like superconducting quantum
computers, the execution of the CNOT gates is subject to
constraints. A physical qubit on the hardware can only interact with
its neighbors, restricting the 2-qubit gates ---such as CNOT--- one
can apply. Taking into account these constraints is a crucial and
difficult task for the design of quantum algorithms and the
optimization of the corresponding quantum circuits. In particular, in
the literature several works present post-processing techniques to
convert with minimum overhead a circuit designed for an ideal hardware
to a circuit designed for a specific architecture
\cite{childs2019circuit}.

Overall the CNOT gate is a costly resource that has to be optimized,
either for NISQ or fault-tolerant computations, and the optimization
may be subject to constraints making the task very challenging.

One way to optimize the CNOT count in quantum circuits is to focus on
circuits consisting solely of CNOT gates, also called linear
reversible circuits. They represent a class of quantum circuits
playing a fundamental role in quantum compilation. They are part of
the so-called Clifford circuits and the CNOT+T circuits, two classes
of circuits that have shown crucial utility in the design of efficient
quantum compilers \cite{2058-9565-4-1-015004,6899791} and error
correcting codes \cite{gottesman1997stabilizer,campbell2012magic}.
For instance the Tpar optimizer \cite{6899791} takes a Clifford+T
circuit as input and decomposes it into a series of CNOT+T circuits
separated by Hadamard gates. Then each CNOT+T circuit is optimized and
re-synthesized by successive syntheses of CNOT circuits and
applications of T gates.

Hence the synthesis of CNOT circuits naturally occurs in general
quantum compilers and giving efficient algorithms for optimizing CNOT
circuits will then be of uttermost importance for both short-term and
long-term applications.

\subsubsection*{Contribution and outline of the paper}

In this paper we focus on the size optimization of linear reversible
circuits. We present a new method for the synthesis of CNOT circuits
relying on solving a well-known cryptographic problem: the syndrome
decoding problem. Our algorithm transforms the synthesis problem into
a series of syndrome decoding problems and we propose several methods
to solve this particular subproblem. This method, initially designed
for a full qubit connectivity, is robust enough to be extended to
partial connectivity.

The outline of the paper is the following: in Section \ref{background}
we present the basic notions and the state of the art in the synthesis
of linear reversible circuits. We first present our algorithm in the
case of an all-to-all connectivity in Section \ref{all}. Then we
extend it to the case of restricted connectivities in Section
\ref{arbitrary}. Benchmarks and discussions are given at the end of Sections \ref{all}
and \ref{arbitrary}. 

This paper is an extended version of a paper published in the
conference proceedings of RC'20.~\cite{RC20}.

\section{Background and state of the art} \label{background}

\paragraph{Synthesis of a linear reversible function}
Let $\mathbb{F}_2$ be the Galois field of two elements.  A linear
reversible function $f$ on $n$ qubits applies a linear Boolean
function on the inputs to each qubit. Given $x \in \mathbb{F}_2^n$ as
inputs, the output of qubit $i$ is
\[
  f_i(x) = \alpha^i \cdot x = \alpha^i_1 x_1 \oplus \alpha^i_2 x_2
  \oplus ... \oplus \alpha^i_n x_n
\]
where $\oplus$ is the bitwise XOR operation and the $\alpha^i$'s are
Boolean vectors also called \textit{parities}. The action of $f$ can
be represented as an $n \times n$ binary matrix $A$ with
$A[i,:] = \alpha^i$ (using Matlab notation for row selection) and
$f(x) = Ax$. In other words each row of $A$ corresponds to the parity
held by the corresponding qubit after application of $A$.  By
reversibility of $f$, $A$ is also invertible in $\mathbb{F}_2$. The
application of two successive operators $A$ and $B$ is equivalent to
the application of the operator product $BA$.

We are interested in synthesizing general linear reversible Boolean
functions into reversible circuits, i.e., series of elementary
reversible gates that can be executed on a suitable hardware. To that
end we use the CNOT gate, it performs the following 2-qubit operation:
\[
  \text{CNOT}(x_1, x_2) = (x_1, x_1 \oplus x_2)
\]
where $x_1$, resp. $x_2$, is the parity held by the control qubit,
resp. the target qubit. If applied after an operator $A$, the total
operator ($A$ + CNOT) is given from $A$ by adding the row of the
control qubit to the row of the target qubit. Such row operations are
enough to reduce any invertible Boolean matrix to the identity matrix,
so the CNOT gate can be solely used to implement any linear reversible
operator. Overall, a CNOT-based circuit can be simulated efficiently:
starting from $A=I$ the identity operator, we read sequentially the
gates in the circuit and apply the corresponding row operations to
$A$.

We use the size of the circuit, i.e., the number of CNOT gates in it
to evaluate the quality of our synthesis. The size of the circuit
gives the total number of instructions the hardware has to perform
during its execution. Due to the presence of noise when executing
every logical gate, it is of interest to have the shortest circuit
possible.

\paragraph{Connectivity constraints}

\begin{figure}[t!]
% \centering
    \begin{subfigure}[b]{0.45\textwidth}
        \centering
        \includegraphics[scale=0.34]{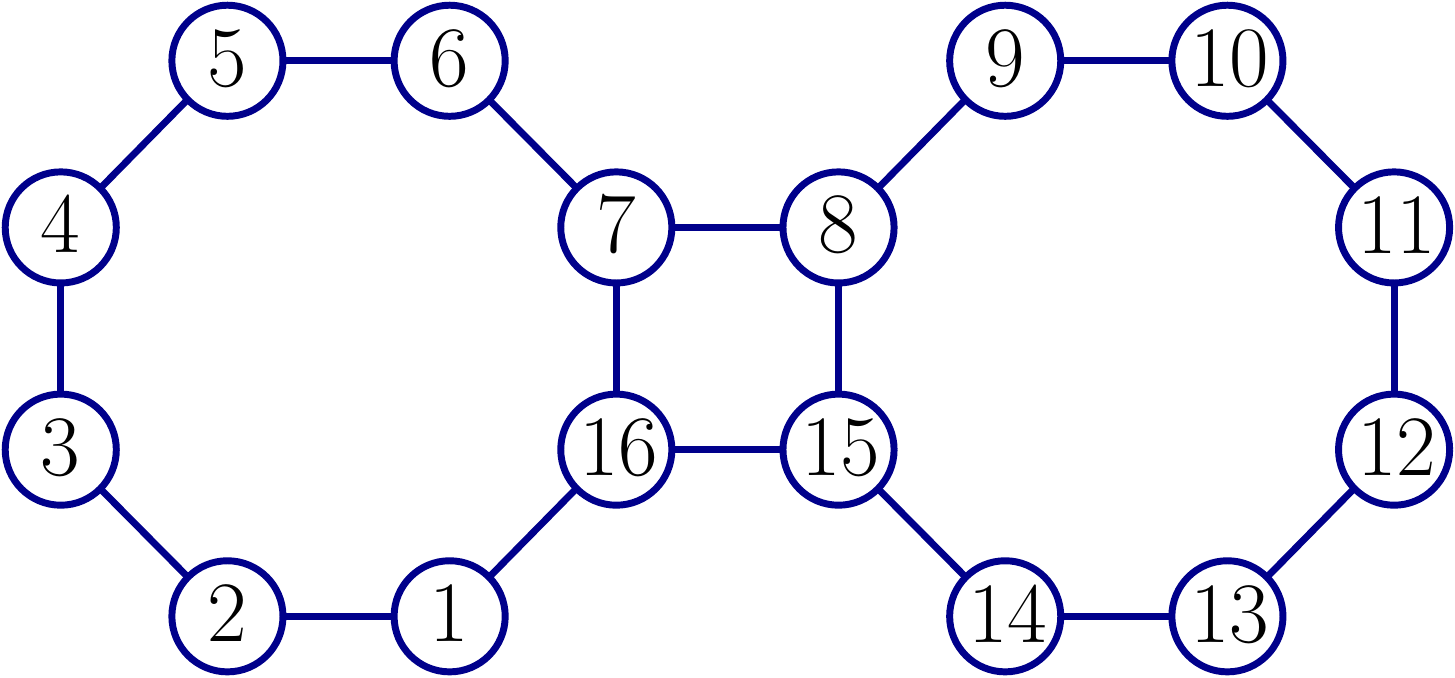}
        \caption{Rigetti 16Q-Aspen}
    \end{subfigure}
    ~
    \begin{subfigure}[b]{0.45\textwidth}
        \centering
        \includegraphics[scale=0.34]{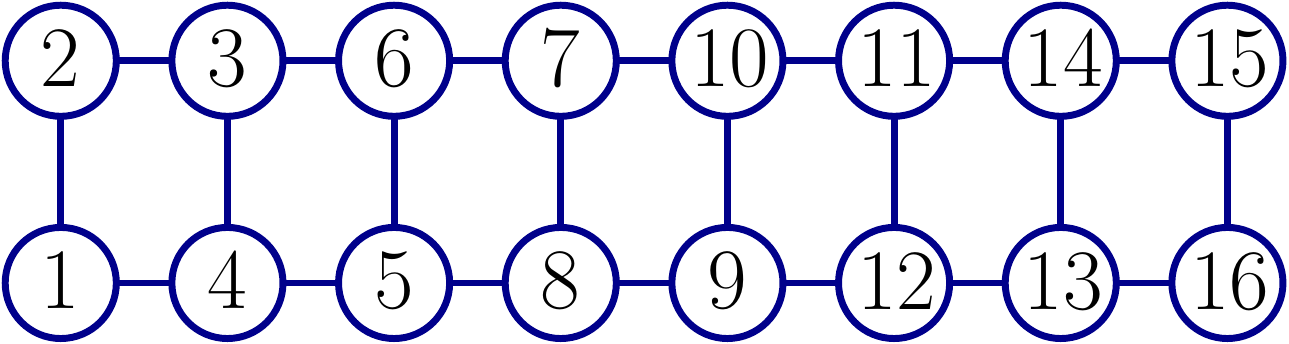}
        \caption{IBM QX5}
    \end{subfigure}

    \begin{subfigure}[b]{0.45\textwidth}
        \centering
        \includegraphics[scale=0.34]{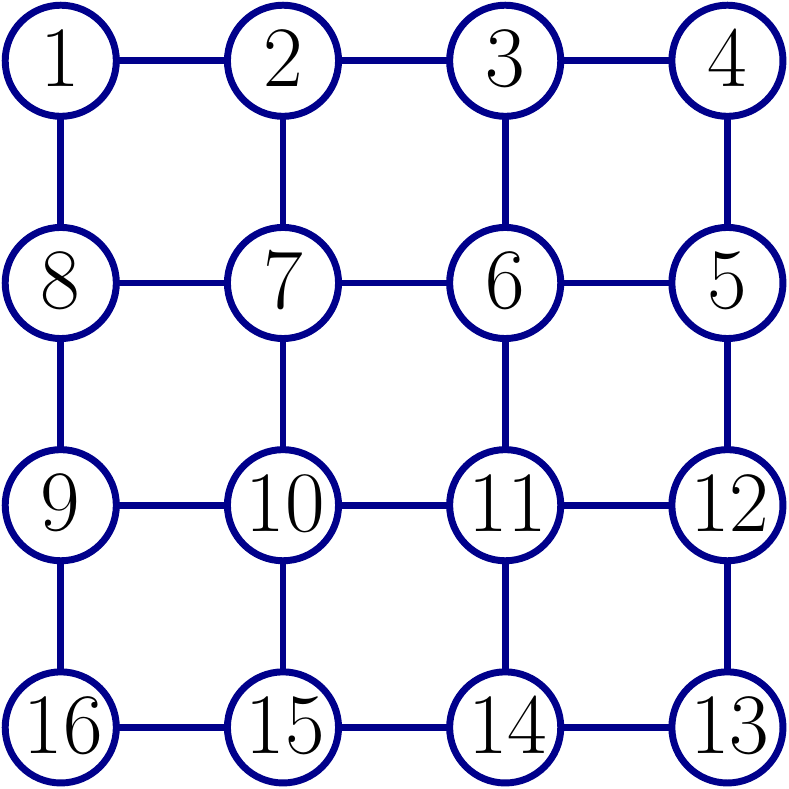}
        \caption{16q-Square}
    \end{subfigure}
    ~
    \begin{subfigure}[b]{0.45\textwidth}
        \centering
        \includegraphics[scale=0.34]{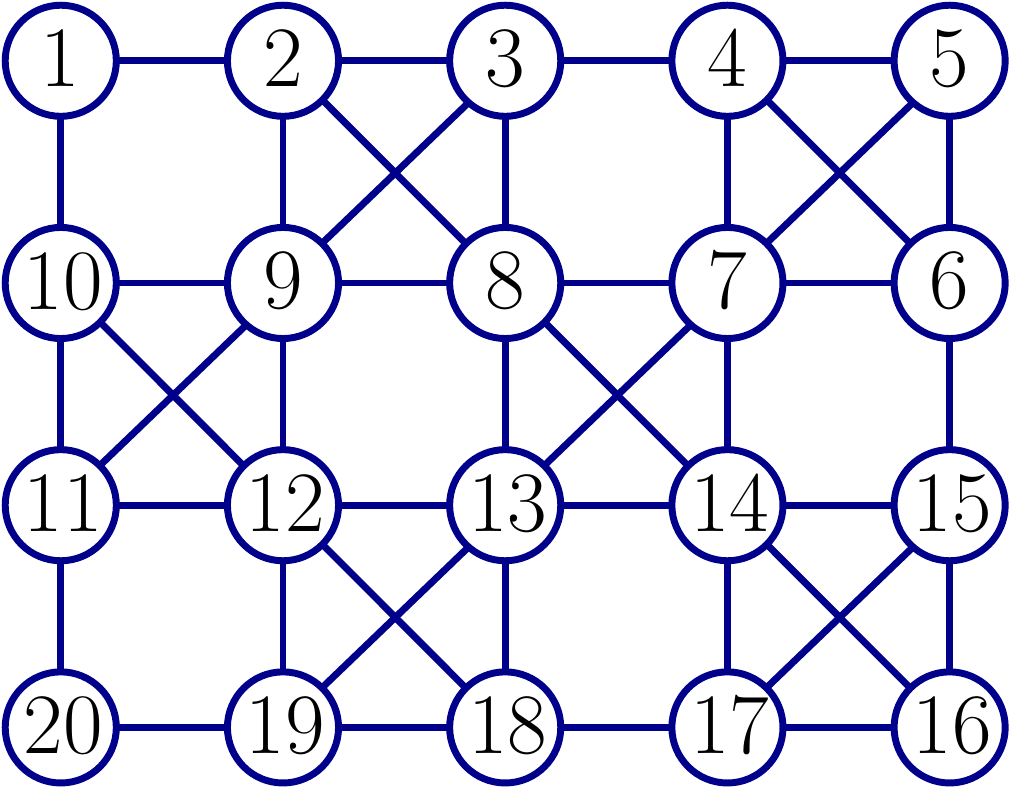}
        \caption{IBM Q20 Tokyo}
    \end{subfigure}
    \caption{Examples of qubit connectivity graphs from existing
      architectures }
\label{architectures}
\end{figure}

At the current time, for superconducting technologies, full
connectivity between the qubits cannot be achieved. The connections
between the qubits are given by a connectivity graph, i.e., an
undirected, unweighted graph where 2-qubit operations, such as the
CNOT gate, can be performed only between neighbors in the
graph. Examples of connectivity graphs from current physical
architectures are given on Fig.~\ref{architectures}.

\paragraph{LU decomposition}
Given the matrix representation $A$ of a generic linear reversible
operator, we can always perform an LU decomposition~\cite{GVL96} such
that there exists an upper (resp. lower) triangular matrix $U$
(resp. $L$) and a permutation matrix $P$ such that $A = PLU$. This
decomposition is not unique, several choices of $(P, L, U)$ are
possible.  The invertibility of $A$ ensures that the diagonal elements
of $L$ and $U$ are all equal to $1$. In the remainder of this paper,
the term ``triangular operator'' stands for an operator whose
corresponding matrix is either upper or lower triangular. The LU
decomposition is at the core of our synthesis of general linear
reversible Boolean operators: synthesizing $U$, $L$, $P$ and
concatenating the circuits gives an implementation of $A$.

\paragraph{State of the art}
In the all-to-all connectivity case the best algorithm is the
Patel-Markov-Hayes (PMH) algorithm {\algoone}. It reaches an
asymptotic optimum and produces circuits of size
$\mathcal{O}(n^2/\log_2(n))$. This algorithm is for instance used in
the Tpar and Gray-Synth algorithms \cite{amy2018controlled,6899791} so
any improvement over {\algoone} will also improve any quantum compiler
that relies on it.

For architectures with restricted connectivity, the first proposed
approach has been to transform the circuits given by an unrestricted
algorithm with swap insertion algorithms to match the connectivity
constraints \cite{li2019tackling,wille2016look,pedram2016layout}. To
produce more efficient circuits, two concomitant papers proposed a
modification of the Gaussian elimination algorithm
\cite{kissinger2020cnot,nash2020quantum}. They synthesize the operator
column by column similarly to the Gaussian Elimination algorithm but
they use Steiner trees to compute the shortest sequence of CNOT gates
for the synthesis of one column.  In \cite{kissinger2020cnot} the
authors compare their method based on Steiner trees against two
compilers: Rigetti Computing’s QuilC and Cambridge Quantum Computing’s
t$\ket{\text{ket}}$ that both produced state of the art results on
benchmarks published by IBM \cite{cowtan2019qubit}. The benchmarks
show a consequent savings in the total number of CNOT gates in favor
of the Steiner tree method, so we consider that the work in
\cite{kissinger2020cnot} is state-of-the-art and we will compare
solely to their algorithm.

For other classes of reversible circuits various methods have been
proposed, see e.g. the use of genetic algorithms for the design of
compressor trees \cite{haghparast2010novel}.

\section{Algorithm for an all-to-all connectivity} \label{all}

In this section we present our algorithm in the case of a complete
connectivity between the qubits. We focus on the synthesis of a lower
triangular operator $L \in F_2^{n \times n}$. What follows can be
straightforwardly extended to the case of upper triangular operators
and to general operators using the LU decomposition. With an
all-to-all connectivity one can avoid to apply the permutation $P$ by
doing a post-processing of the circuit that would transfer the
permutation operation directly at the end of the total circuit. This
can be done without any overhead in the number of gates.

A circuit implementing $L$ can solely consist of ``oriented'' CNOTs,
whose controlled qubit $i$ and target qubit $j$ satisfy $i < j$. The
circuit given by the Gaussian elimination algorithm is an example. For
this particular kind of circuits, a CNOT applied to a qubit $k$ does
not have any influence on the operations performed on the first $k-1$
qubits: removing such a CNOT will not modify the result of the
synthesis of the first $k-1$ parities. We use this property to design
a new algorithm where we synthesize $L$ parity by parity and where we
reuse all the information acquired during the synthesis of the first
$k$ parities to synthesize parity $k+1$.

Given $L_{n-1} = L[$1:$n-1$, 1:$n-1]$ (again using Matlab notation), a
circuit $C$ implementing the operator
$\left(\begin{smallmatrix} L_{n-1} & 0 \\ 0 &
    1 \end{smallmatrix}\right)$ and considering that we want to
synthesize the operator
$L = \left(\begin{smallmatrix} L_{n-1} & 0 \\ s &
    1 \end{smallmatrix}\right)$ the core of our algorithm consists in
adding a sequence of CNOTs to $C$ such that we also synthesize the
parity $s$ of the $n$-th qubit. During the execution of $C$, applying
a CNOT $i \to n$ will add the parity currently held by qubit $i$ to
the parity of qubit $n$ without impacting the synthesis of the first
$n-1$ parities. In other words, if we store in memory all the parities
that appeared on all $n-1$ qubits during the execution of the circuit
$C$, we want to find the smallest subset of parities such that their
sum is equal to $s$. Then when a parity belonging to this subset
appears during the execution of $C$, on qubit $i$ for instance, we
insert in $C$ a CNOT $i \to n$. We ultimately have a new circuit $C'$
that implements $L$.

The problem of finding the smallest subset of parities whose sum
equals $s$ can be recast as a classical cryptographic
problem. Assuming that $H \in F_2^{n-1 \times m}$ is a Boolean matrix
whose columns correspond to the $m$ available parities, any Boolean
vector $x$ satisfying $Hx = s^T$ gives a solution to our problem and
the Hamming weight of $x$, $wt(x)$, gives the number of parities to
add, i.e., the number of CNOTs to add to $C$. We are therefore
interested in an optimal solution of the problem
\begin{equation} 
\begin{aligned} 
& \underset{x \in F_2^m}{\text{minimize}} & wt(x) \\
& \text{such that} & Hx = s^T.
\end{aligned}
\label{syndrome}
\end{equation}

Problem \ref{syndrome} is an instance of the {\it syndrome decoding
  problem}, a well-known problem in cryptography. The link between
CNOT circuit synthesis and the syndrome decoding problem has already
been established in \cite{amy2018controlled}, yet it was used in a
different problem for proving complexity results (under the name of
Maximum Likelihood Decoding problem) and the authors did not pursue
the optimization.  The syndrome decoding problem is presented in more
details in Section~\ref{sec:syndrome}.

To summarize, we propose the following algorithm to synthesize a
triangular operator $L$. Starting from an empty circuit $C$, for $i$
from $1$ to $n$ perform the three following steps:
\begin{enumerate}
\item scan circuit $C$ to compute all the parities available on a
  single matrix $H$,
\item solve the syndrome decoding problem $Hx = s$ with $s$ the parity
  of qubit~$i$,
\item add the relevant CNOT gates to $C$ depending on the solution obtained.
\end{enumerate}
Provided that the size of $C$ remains polynomial in $n$, which will be
the case, then steps 1 and 3 can be performed in polynomial time and
in practice in a very short amount of time. The core of the algorithm,
both in terms of computational complexity and final circuit
complexity, lies in Step 2.

\subsection{Syndrome decoding problem}
\label{sec:syndrome}

In its general form, the syndrome decoding problem is known to be
NP-Hard \cite{berlekamp1978inherent} and cannot be approximated by a
constant factor \cite{arora1997hardness}.
A good overview of how difficult the problem is can be found in
\cite{vardy1997algorithmic}. 

We give two methods for solving the syndrome decoding problem. The
first one is an optimal one and uses integer programming solvers. The
second one is a greedy heuristic for providing sub-optimal
results in a short amount of time.

\subsubsection{Integer programming formulation}

The equality $Hx = s$ is a Boolean equality of $n$ lines. For instance
the first line corresponds to
\[ H_{1,1} x_1 \oplus H_{1,2} x_2 \oplus \hdots \oplus H_{1,m} x_m =
  s_1. \]
We transform it into an ``integer-like'' equality constraint. A
standard way to do it is to add an integer variable $t$ and to create
the constraint
\[ H_{1,1} x_1 + H_{1,2} x_2 + \hdots + H_{1,m} x_m  - 2t = s_1. \]
If we write $c = (1,...,1,0,...,0)^T \in \mathbb{N}^{m+n}$ and
$A = [H | -2I_n]$ then the syndrome decoding problem is equivalent to
the integer linear programming problem
\begin{equation} 
\begin{aligned} 
& \underset{x \in F_2^{m}, t \in \mathbb{N}^n}{\min} & c^T \cdot [x;t] \\
& \text{such that} & A[x ; t] = s.
\end{aligned}
\end{equation}

\subsubsection{A cost minimization heuristic} \label{sec::cost_heuristic}

Although the integer programming approach gives optimal results, it is
very unlikely that it will scale up to a large number of
qubits. Moreover, to our knowledge the other existing algorithms
proposed in the literature give exact results, they are complex to
implement and their time complexity remains exponential with the size
of the problem. We therefore have to consider heuristics to compute an
approximate solution in a much shorter amount of time.

We use a simple cost minimization approach: starting with the parity
$s$ we choose at each iteration the parity $v$ in $H$ that minimizes
the Hamming weight of $v \oplus s$ and we pursue the algorithm with
the new parity $v \oplus s$. The presence of the canonical vectors in
$H$ (as we start with the identity operator) is essential because they
ensure that this method will ultimately converge to a solution.

A simple way to improve our heuristic is to mimic
path finding algorithms like Real-Time A*
\cite{korf1990real}. Instead of directly choosing the parity that
minimizes the Hamming weight, we look up to a certain horizon and we
make one step in the direction of the most promising path.
To control the combinatorial explosion of the number of paths to
browse, we only expand the most promising parities at each level. We
set the maximum width to $m$ and the depth to $k$ so that it
represents at most $m^k$ paths to explore. With suitable values of $m$
and $k$ we can control the total complexity of the algorithm. A
limitation of such a simple approach is that we can store the same
path but with different parities order: we decided to ignore this
limitation in order to keep a simple implementation.

Lastly, we introduce some randomness by solving
the problem $PHx = Ps$ for several random change of basis matrices
$P$. Repeating this several times for one syndrome decoding problem
increases the chance to find an efficient solution. This technique has
been proven to be efficient for a class of cryptographic algorithms
called Information Set Decoding \cite{prange1962use}, even though the
complexity of these algorithms remains exponential.

\subsection{Benchmarks} \label{sec::benchs}

All the code is written in Julia and executed on a MacBook Air 1.8 GHz
Intel Core i5.

We generate random operators by generating random circuits with
randomly placed CNOT gates. When the number of input gates is
sufficiently large we empirically note that the operators generated
represent the worst case scenario.

We first generate an average complexity for different problem sizes:
for $n=1..150$ we generated 20 random operators on $n$ qubits with
more than $n^2$ gates to reach with high probability the worst
cases. 

We present our results in three batches, one for each class of methods
used for solving the syndrome decoding problem:
\begin{itemize}
  \item the cost minimization methods,
  \item the simplest cost minimization method with random changes of
    basis (the Information Set Decoding strategy),
  \item the integer programming solver.
\end{itemize}

% We run our algorithms on this set of operators in the following
% cases:
% \begin{itemize}
% 	\item with the integer programming solver (Coin-or branch and cut solver),
% 	\item with the cost minimization heuristic with unlimited width and depth 1,
% 	\item with the cost minimization heuristic with width 60 and depth 2,
% 	\item with the cost minimization heuristic with width 15 and depth 3,
% 	\item with the cost minimization heuristic (width=Inf, depth=1) and $50$ random changes of basis, the ``ISD'' case.
% \end{itemize}

% In the case of the ISD experiment, due to its probabilistic
% nature, one can hope that repeating the complete synthesis several
% times and keeping the shortest circuit would improve the results. Yet
% the experiments show that it has a minor influence on the final
% result.

\subsubsection{Greedy solvers comparison.}

We first present the results of the cost minimization techniques. We
compare the performance of the greedy methods for different values of
the width and depth search. For a fair comparison, the size of the
search tree must be roughly the same in each case, i.e., at
$\text{width}^{\text{depth}}$ fixed.  We benchmarked the following
cases:
\begin{itemize}
  \item width=Inf, depth=1,
  \item width=60, depth=2,
  \item width=15, depth=3,
  \item width=8, depth=4,
  \item width=5, depth=5.
\end{itemize}

\begin{figure}[ht]
\centering
\includegraphics[scale=0.5]{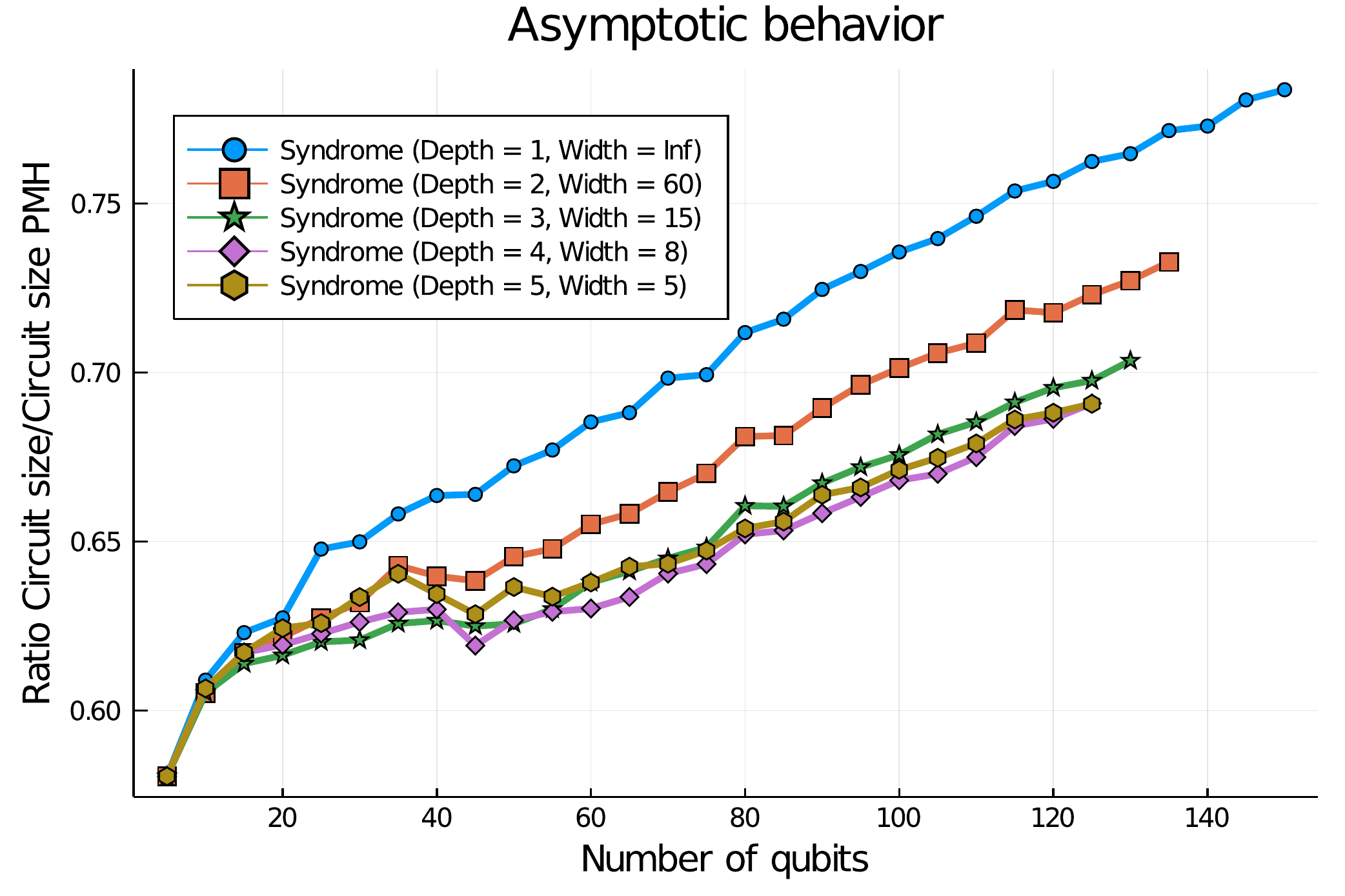}
\caption{Average performance of the Syndrome Decoding based algorithms
  with greedy solvers versus the state of the art {\algoone}.}
\label{asymptotic_graph_greedy}
\end{figure}

The results are given in Fig.~\ref{asymptotic_graph_greedy}. For
clarity, instead of plotting the size of the circuits we plot the
ratio between the size of the circuits given by our algorithms and the
state of the art algorithm {\algoone}. We stopped the calculations
when the running time was too large for producing benchmarks in
several hours.

Overall, it seems better to increase the depth of search than the
width. Yet, for a depth of 3 and more, the improvements are not as
clear.  Clearly the worst results are for the depth 1 and depth 2
cases. Then it seems that the decreasing of the width plays a role as
the results for depth 5 are not as consistent as for depth 4. This
might be due to the fact that we store several times the same
solutions: with small width we increase the chances to store only a
few different solutions but with several vector orders.

For problems smaller to $130$ qubits we manage to outperform
{\algoone} with more than $30\%$ of gain. Then, as the number of
qubits increases our method performs worse. We ran a few computations
for much larger problems and the results are that {\algoone} produces
shorter circuits whenever $n$ goes approximately beyond $400$. This
raises the question of whether it is due to the method in itself or to
the solution of the syndrome decoding that becomes less and less
optimal as the problem size increases. We leave this question as a
future work.

\subsubsection{Greedy method + Information Set Decoding (ISD)}

We now investigate the role of doing the cost minimization process but
with repeated random changes of basis. For computational reasons, we
can only repeat the case depth=1, the fastest by far, as repeating the
others would take too much time. To investigate how the performance of
the algorithm varies with the number of iterations, we considered the
following cases:
\begin{itemize}
  \item with 50 iterations,
  \item with 100 iterations,
  \item with 500 iterations,
  \item with 1000 iterations,
  \item with 10000 iterations.
\end{itemize}

\begin{figure}
  \centering
  \includegraphics[scale=0.5]{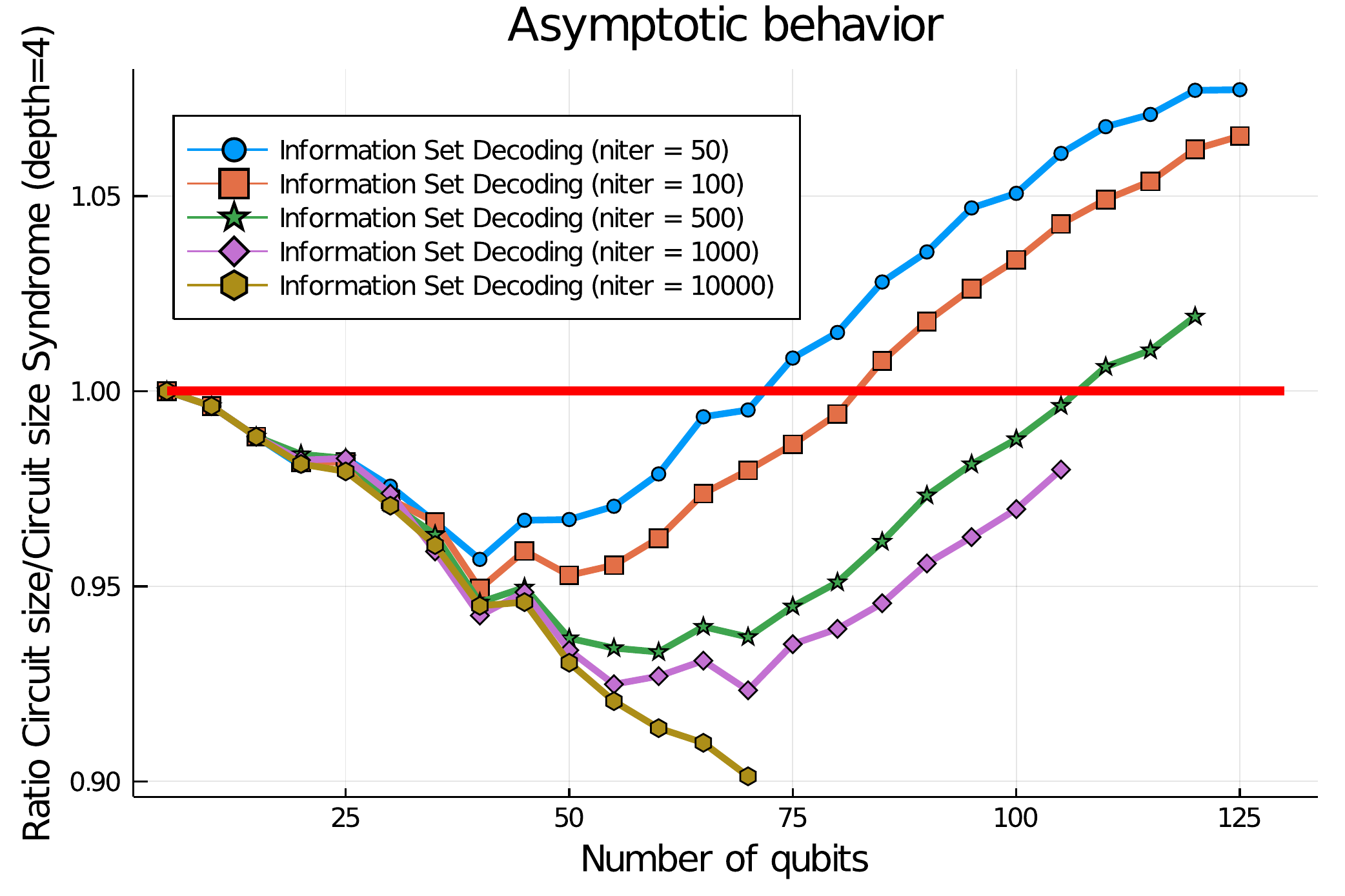}
  \caption{Average performance of the ISD strategy against the cost
    minimization technique with depth=4.}
\label{asymptotic_graph_isd}
\end{figure}

The results are given in Fig~\ref{asymptotic_graph_isd}. We focused on
the range $n=1...130$. Given that the greedy solver with depth=4
approximately provides the best results from the greedy solvers
category, and overall the best results, we plot the ratio between the
circuit size returned by the ISD strategies and the circuit size
returned by this new state-of-the-art method.

Not surprisingly, increasing the number of iterations increases the
chances to find good solutions to the syndrome decoding problem, thus
to find short circuits. What is more surprising maybe is the
efficiency of such process. In terms of computational time, the case
niter=500 is roughly as costly as using a greedy solver with depth
4. So for an equivalent amount of computational time, the ISD strategy
outperforms our greedy solvers with up to 8\% of gain. With more
iterations we reach up to 10\% of savings.

Overall, the ISD strategy works well until the search space is too
large to be efficiently explored with a random method and so few
tries. Inevitably, even with niter=1000, the performance of the ISD
strategy deteriorates and eventually the performance of the solver is
close to the initial greedy solver with only one try. Nonetheless, we
believe this gives good insights on the question we asked above,
whether the bad performances of the syndrome decoding based methods on
large instances are due to the method in itself or in the quality of
the solutions of the syndrome decoding problems. When we increase the
number of iterations we significantly improve the performance of the
method and the range of validity of the method. It seems to go in the
sense that it is difficult to optimally solve the syndrome decoding
problem and that this has direct consequences on the final result.

\subsubsection{Integer programming solvers comparison.}

We are able to use an integer programming solver for problems up to 50
qubits.  For larger problems, the exponential scalability of such
solvers make it impossible to use them in a reasonable amount of
time. Nonetheless, it is possible to use integer programming solvers
to "improve" a solution. Indeed, for some solvers we can ask them to
focus on a strategy that tries to find quickly a good solution. Then
if we force the solver to stop after a certain time limit, we can
control the overall complexity of our syndrome decoding solver. Given
that we can give the solver an initial solution to the problem, we are
ensured that the solver will return a valid solution.

Unfortunately, the results are not significant. For problems on 50
qubits or less, our results are identical to the ones obtained with
the ISD strategy, proving that this method can almost optimally solve
the syndrome decoding problem on small instances. For larger problems,
we gave the solver the solution of the ISD strategy with $500$
iterations and we let the solver compute a better solution during 5
seconds. Given that for an operator on $n$ qubits we have to solve
$2n$ instances of the syndrome decoding problem, this represents a non
negligible amount of extra time. Overall, the savings are negligible
($<1\%$) compared to the ISD strategy with $500$
iterations. Increasing the number of iterations in the ISD strategy
gives much better savings and in a much shorter time, so overall
integer programming solvers should not be the first option to use.

\subsubsection{Experiments on simpler operators}

\begin{figure}[h]
\centering
\includegraphics[scale=0.5]{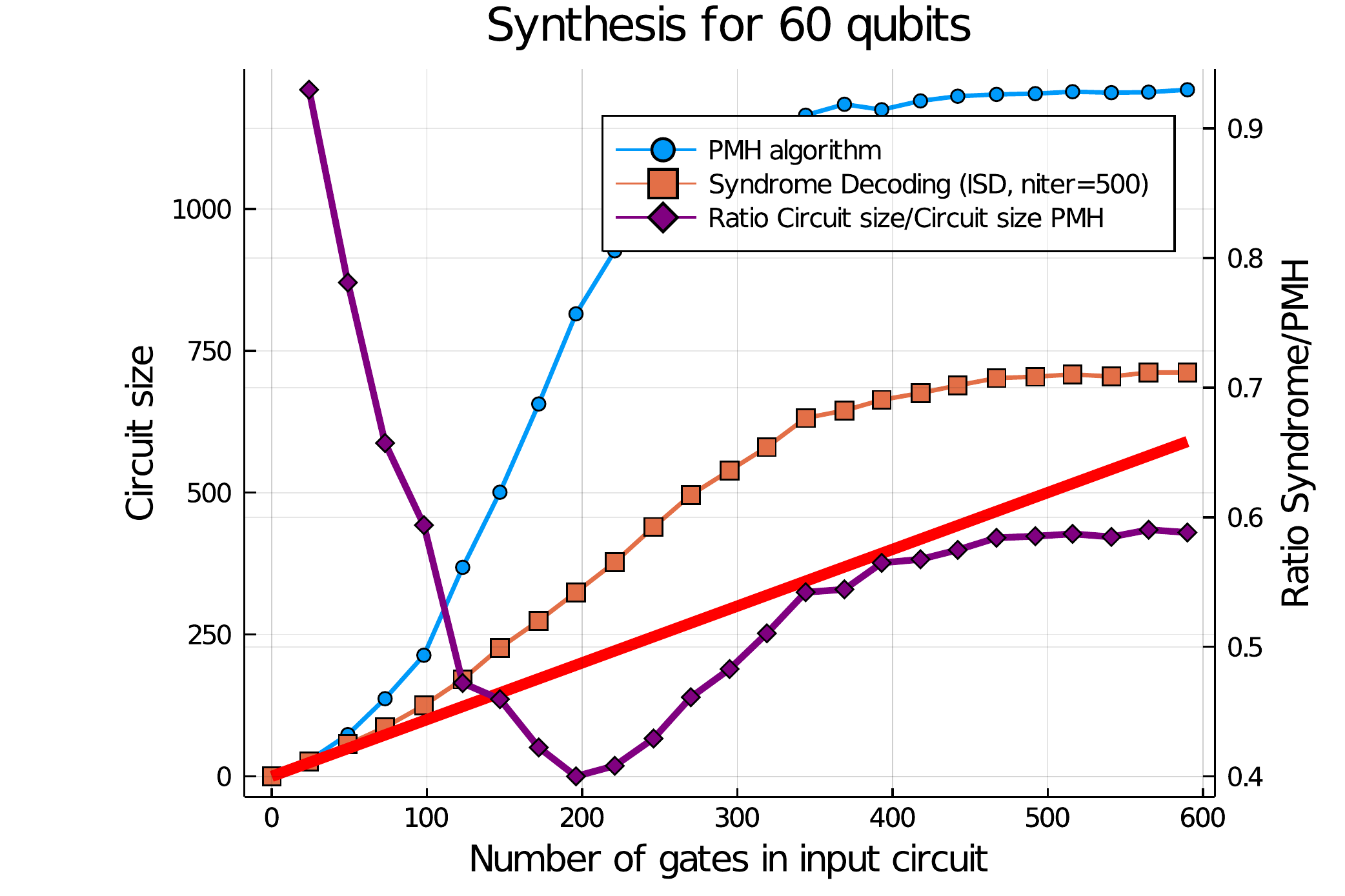}
\caption{Performance of the syndrome decoding based algorithm versus {\algoone} on 60 qubits for different input circuit sizes.}
\label{60qubit_graph}
\end{figure}

We now look at the performance of the algorithms on a
specific number of qubits, here $n=60$, but for different input
circuit sizes. This experiment reveals how close to optimal our
algorithm is when we synthesize an operator for which we expect a
small output circuit. The results are given in
Fig.~\ref{60qubit_graph}. As the ISD method produces the best results
for this size of problem we only plot the results for this method. We
also plot the line $y=x$ that shows how far we still are from the
optimal solution. Again we outperform the best algorithm in the
literature even for small input circuits with more than 50\% of
savings when the input circuit is of size 100-300 gates, with a
maximum saving of 60\% for approximately 200 gates.

\subsection{Discussion}

We now propose some open questions on how to improve the syndrome
decoding based algorithm.

\subsubsection{Theoretical complexity}

Due to the use of heuristics, we cannot give a tight worst-case
complexity of our algorithm in terms of circuit size. Notably, we
cannot prove that the circuits will be of size $O(n^2/\log_2(n))$ in
the worst case, as done in \cite{patel2008optimal}. We can only
guarantee a complexity in $O(n^2)$. It would be interesting as a
future work to use cryptographic tools in order to prove theoretical
guarantees, even in more restricted cases (for instance if we can
solve the syndrome decoding problem exactly).

\subsubsection{A global solver}

In the benchmarks presented in Section~\ref{sec::benchs}, we
highlighted a strong correlation between the quality of a solver for
the syndrome decoding problem and the quality of the overall CNOT
synthesis algorithm. Although such correlation seems quite intuitive,
it does not answer the following question: does solving optimally the
syndrome decoding problem provide an optimal synthesis algorithm for a
triangular operator?

There may be cases where making some parities appearing on instances
of the syndrome decoding problem will help solve following syndrome
decoding problems. If this happens to be true, designing a global
version of the syndrome decoding problem would be a challenging
task. Notably the computational complexity of such algorithm could be
intractable. Can we design a cheap solver for the syndrome decoding
algorithm taking into account other future syndrome decoding problems?

\subsubsection{An alternative formulation of the syndrome decoding problem} \label{sec::crypto}

\paragraph{Navigating through the space of solutions with a generator
  matrix} We propose an alternative formulation of the syndrome
decoding problem using the properties of linear codes. A linear code
$C$ of length $n$ and rank $k$, noted $[n,k]$, is a $k$-dimensional
subspace of the vector space $\mathbb{F}_2^n$. A linear code is
characterized notably by two matrices: the generator matrix
$G \in \mathbb{F}_2^{n \times k}$ and the parity-check matrix
$H \in \mathbb{F}_2^{(n-k) \times n}$. Any codeword $y$ in $C$ is
generated by $G$, i.e., $y = Gx$ for some $x$, and for any codeword
$y$ we also have $Hy = 0$. When sending a codeword $y$ through a noisy
channel, we may recover an altered word $z$ such that $z = y+e$ does
not belong to $C$ anymore. $e$ is the error done during the
transmission and the decoding process consists in finding which $y$ in
$C$ is the closest to $z$. Equivalently we want to find $e$ with the
minimum Hamming weight. Applying $H$ to the received word and we have
\[ Hz = H(y+e) = Hy + He = He. \] Setting $s = Hz$ and we recover the
\textit{syndrome decoding problem}. In this case any solution of the
syndrome decoding problem can be written
\[ \{ x_0 + Gx \; | \; x \in \mathbb{F}_2^k \} \] where $x_0$ is any
solution of the problem. In some sense, the generator matrix $G$ is
related to the pseudo-inverse of $H$. There is a simple way to compute
$G$ from $H$ and vice-versa: if
$G = \begin{pmatrix} I_k \\ P \end{pmatrix}$ for some matrix $P$ then
$H = \begin{pmatrix} I_{n-k} & P^T \end{pmatrix}$ and reciprocally.

In our case, the length of our linear code is given by the number of
parities available $m$ and the dimension of our code is $k=m-n$ ($n$
is the number of qubits on which the parities are encoded). One can
easily check that $H \in \mathbb{F}_2^{m-k \times k}$. Given that we
always have $H = \begin{pmatrix} I_{m-k} & P \end{pmatrix}$ (the
canonical vectors are always in $H$ as the first parities available)
then the computation of $G$ is straightforward.  Contrary to our
initial formulation of the syndrome decoding problem, we can navigate
directly in the space of the solutions of the problem in order to find
the best one. Can we exploit this to have a more efficient algorithm
for the syndrome decoding problem?

\begin{figure}
\center
\includegraphics[scale=0.8]{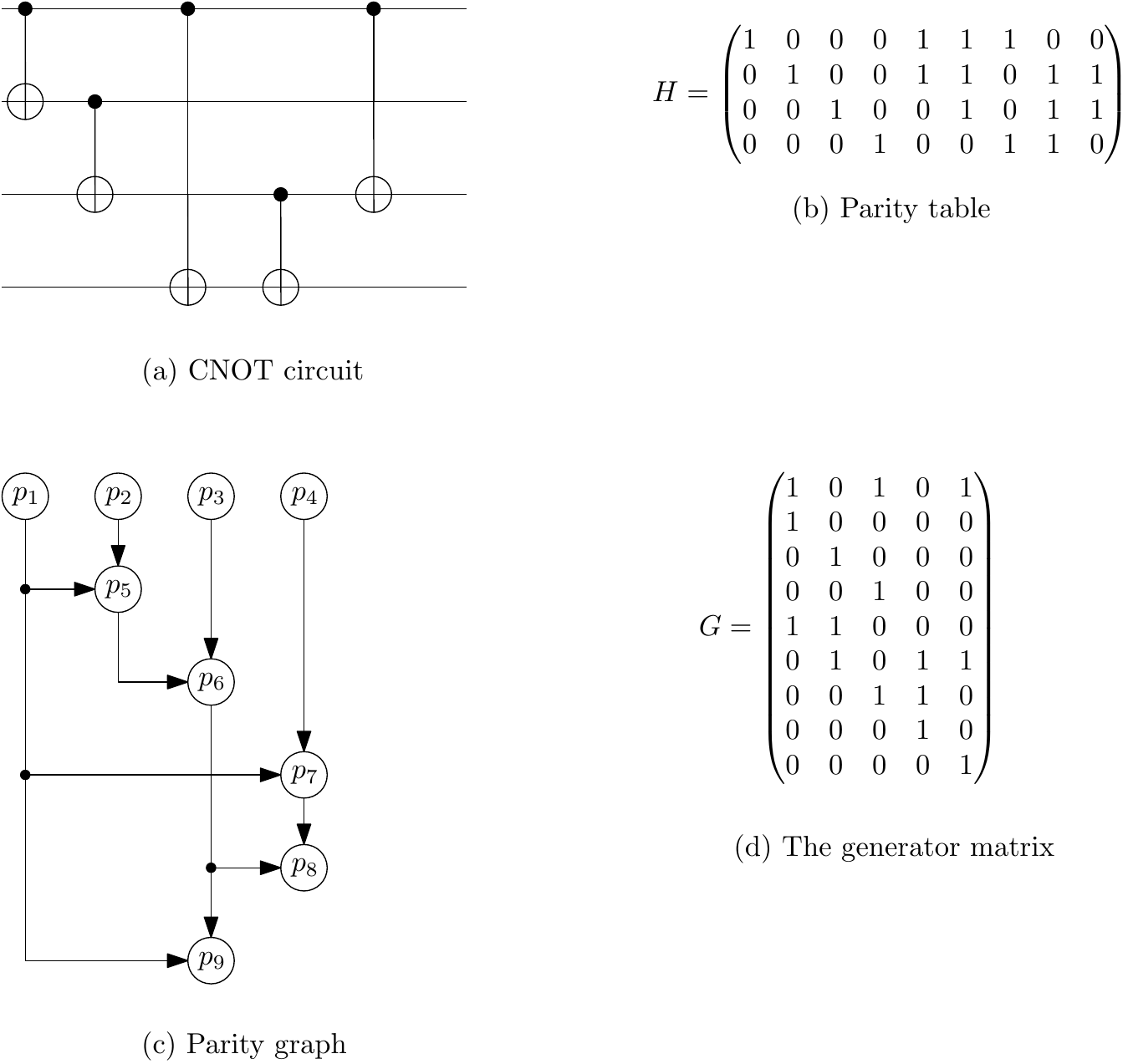}
\caption{An example of a CNOT circuit, the parities that appear in it, the associated parity graph and the generator matrix $G$ associated to the linear code.}
\label{cnot::parity_graph}
\end{figure}

\paragraph{A graph-oriented formulation} In this paper, we give a
complementary graph-oriented approach to this problem. This will
provide new understandings on the structure of the problem and notably
this will show a new way to compute the generator matrix $G$. When we
scan a circuit to compute the available parities, each new parity is
obtained because we apply a CNOT gate that sums two rows of the
current linear reversible operator. So, when we gather the different
parities in a matrix $H$, if the gathering is done chronologically,
each new column of $H$ can be written as a sum of two previous columns
of $H$. We can create a parity graph where each node is a parity and,
given three nodes $v_1, v_2, v_3$ corresponding to three parities
$p_1, p_2, p_3$, we add two edges $v_1 \to v_3, v_2 \to v_3$ if
$p_3 = p_1 \oplus p_2$. We call such a three-node structure a
triangle. Each node except the first $n$ ones have two in-edges. There
is no restriction on the number of out-edges though. A solution to the
syndrome decoding problem is given by a subset of nodes: those nodes
are considered active. Given such a subset, we can switch to another
solution by considering one triangle and taking the complementary in
terms of active/inactive nodes. We can then navigate in the space of
all the solutions in search of the one with the fewest active nodes:
this is the optimal solution of the syndrome decoding problem. One can
show that we can always reach the optimal solution with only the
elementary operation "take the complementary of a triangle" because
whatever the set of active nodes, we can always reach a canonical form
where the active nodes are solely among the first $n$ nodes.  This
problem gives a new formulation of the $G$ matrix where each column
has only three nonzero elements. The problem is illustrated on $4$
qubits in Fig~\ref{cnot::parity_graph}. Does there exist an efficient
way to solve it?

\section{Extension to an arbitrary connectivity} \label{arbitrary}

In this section we extend the algorithm to the case where the
connectivity is not complete.  First we present how to adapt our
algorithm based on the syndrome decoding for the synthesis of
triangular operators, then we extend our method to the synthesis of
any general operator.

\subsection{Synthesis of a triangular operator}

Let $G$ be a qubit connectivity graph and $L$ the lower triangular
operator to synthesize. We require an ordering on the nodes of $G$
such that the subgraphs containing only the first $k$ nodes, for
$k=1..n$, are connected. As we need to synthesize both $L$ and $U$ we
need in fact this property to be true for an ordering of the qubits
and the reverse ordering. An Hamiltonian path in $G$ is enough to have
this property so for simplicity we assume that the ordering follows an
Hamiltonian path in $G$.

\begin{figure}
\begin{center}
\includegraphics[scale=0.75]{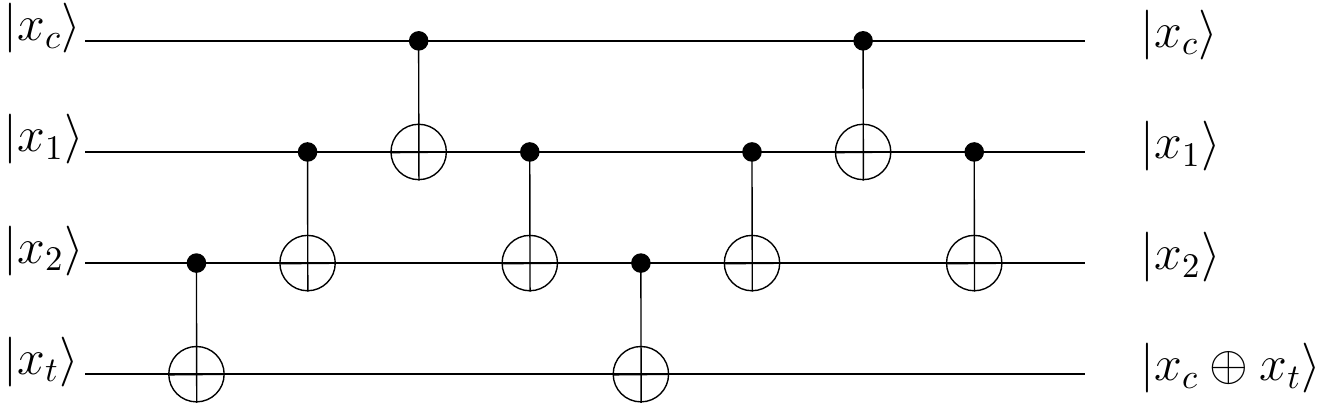}
\end{center}
\caption{CNOT in LNN architecture}
\label{CNOT_LNN}
\end{figure}

Even though the native CNOTs in the hardware are CNOTs between
neighbor qubits in the connectivity graph, it is possible to perform
an arbitrary CNOT gate but this requires more local CNOT gates. Given
a target qubit $q_t$ and a qubit control $q_c$ and assuming we have a
path $(q_c, q_1, ..., q_k, q_t)$ in the graph connecting the two nodes
(such path always exists with the assumption we made above), it is
possible to perform the CNOT $q_c \to q_t$ with $\max(1,4k)$ CNOTs. An
example for 4 qubits (with $k=2$) is given in Fig~\ref{CNOT_LNN}.

Hence, it is still possible to perform the synthesis parity by parity
but we have to be more careful in the setting and in the solution of
the syndrome decoding problem. Not all parities have the same cost,
depending on the qubit holding the parity and its position on the
hardware.

Therefore we have to solve a weighted version of the syndrome decoding
problem. Namely once we have a set of parities in a matrix $H$ and a
cost vector $c \in \mathbb{N}^m$, we look for the solution of the
optimization problem
\begin{equation} 
\begin{aligned} 
& \underset{x \in F_2^m}{\text{minimize}} & c^T \cdot x \\
& \text{such that} & Hx = s^T.
\end{aligned}
\label{weighted_syndrome}
\end{equation}

Problem \ref{weighted_syndrome} can be recasted again as an integer
linear programming problem: we only have to change the value of
$c$. We also propose a greedy heuristic for solving quickly and
approximately the problem: we define the ``basis cost'' of
implementing $s$ as the sum of the costs of each canonical vector
whose component in $s$ is nonzero. Let $\text{bc}(s)$ be this
cost. Our greedy approach consists in finding among the parities of
$H$ the parity $v$ (column $i$ of $H$) that minimizes the cost
\[ c[i] + \text{bc}(s \oplus v). \]

This approach gives a good trade-off between zeroing the most costly
components of $s$ and applying parities at a very high cost. Again we
can repeat the algorithm with random changes of basis to find a better
solution. Especially we focused on computing bases for which the
canonical vectors have the lowest possible costs. In other words, the
new canonical vectors correspond to parities hold by qubits that are
as close as possible to the target qubit.

Nonetheless, compared to the all-to-all case, solving the weighted
syndrome decoding problem is not the only computational core for
controlling both the quality of the solution and the computational
time. Another key task lies in the enumeration of the available
parities. As we will see, it is possible to generate more parities for
one syndrome decoding problem instance and this increases the chances
to get a low-cost solution.

\subsubsection{Listing the parities available.} Until now we set the
weighted syndrome decoding instances by computing the parities
appearing during the synthesis and by using the template in
Fig.~\ref{CNOT_LNN} to estimate their costs. This is in fact
inefficient because it ignores some specificities of the problem:
\begin{itemize}
\item It is possible to add multiple parities in one shot using the
  template in Fig.~\ref{CNOT_LNN}.
\item There is not necessarily one unique path in $G$ between the
  control qubit and the target qubit.
\end{itemize}

\begin{figure}
\begin{center}
\includegraphics[scale=0.59]{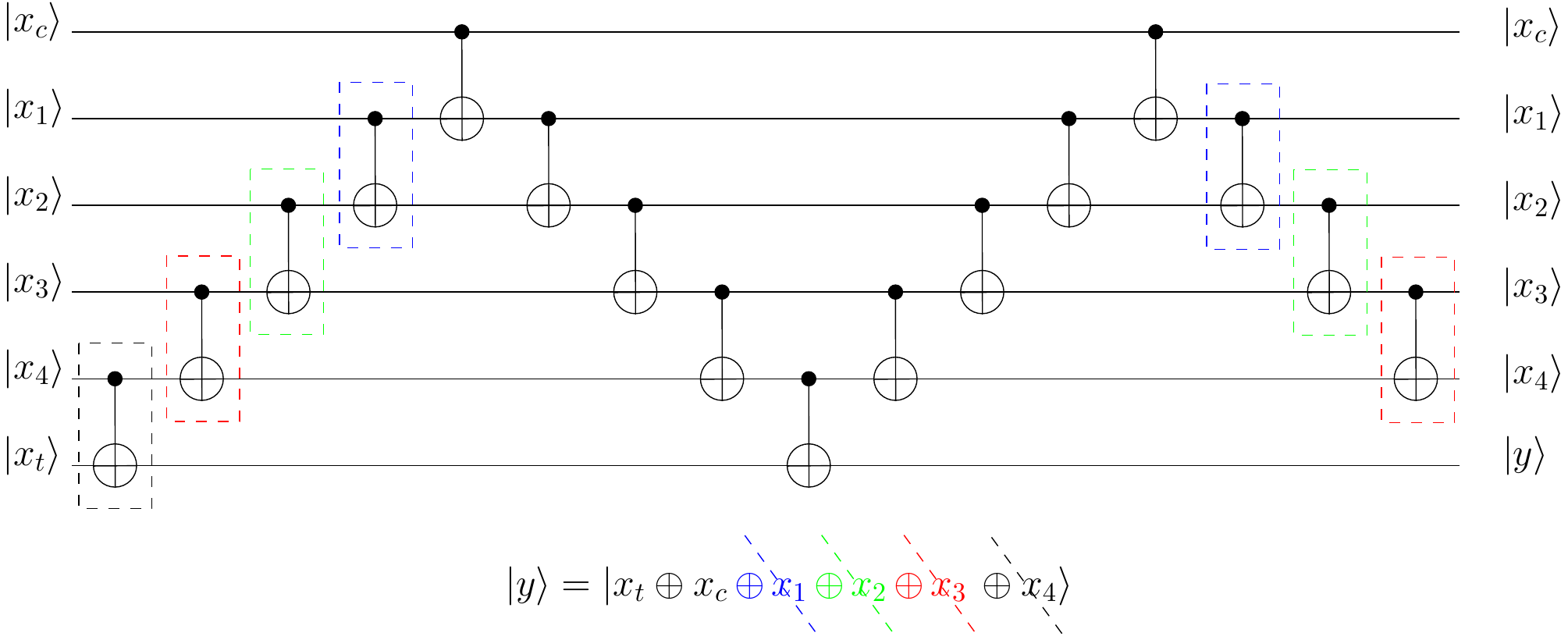}
\end{center}
\caption{Fan in CNOT in LNN architecture}
\label{multi_CNOT_LNN}
\end{figure} 

More precisely, the template shown in Fig.~\ref{CNOT_LNN} is the best
to our knowledge, in terms of size, to apply solely the parity on
qubit $q_c$ to qubit $q_t$. However it is possible to apply any parity
\begin{equation}
  q_t \leftarrow q_t \oplus q_c \oplus_{i=1}^k \alpha_i
  q_i \label{eq::lc}
\end{equation}
with $\alpha_i \in \{0,1\}$ using less CNOTs than required for
applying only $q_c$. In fact the less costly linear combination of
parities is the complete combination
$q_c \oplus q_1 \oplus ... \oplus q_k$, for which $2k+1$ CNOTs are
enough. Removing any parity from this combination requires 2
additional CNOTs per parity except for the qubit $q_k$ that needs only
one extra CNOT. An explanatory template on 6 qubits ($k=4$) is given
on Fig.~\ref{multi_CNOT_LNN}. For any parity at a distance $k$ of the
target qubit, there is at most $2^{k-1}$ different linear combinations
possible and just as many new parities to consider. Moreover the path
between the control qubit and the target qubit matters as a different
path will result in different linear combinations of parities. A
slight modification of the A* algorithm is enough to compute all the
shortest paths between two nodes in a graph.

Even for a small number of qubits the number of parities becomes
quickly intractable. The number of linear combinations along a path
increases exponentially with the length of the path as the number of
paths for most of the architectures --- a grid for instance. In
practice we control the total number of parities by favoring paths
over the choices in the linear combinations. This option is
empirically justified but a more detailed analysis could be made. For
one path we only consider the less costly linear combination, i.e.,
the one that adds all the parities on the way. On the other hand if
possible we go through all the shortest paths between one control
qubit and one target qubit.

\subsubsection{A faster heuristic}

For large problem sizes, even with the minimum number of paths and
linear combinations possible, the number of parities can quickly
become intractable because one CNOT gate can still introduce up to
$n-2$ new parities. Indeed, in a straight line for instance given by a
path $q_1 \to q_2 \to \hdots \to q_n$, modifying qubit $q_{n-1}$ will
add all possible linear combinations $ \sum_{i=k}^{n-1} q_i$ for
$k=1...n-1$. This potential factor of $n$ quickly becomes a non
negligible computational overhead.

We propose therefore a simpler but faster heuristic. Instead of
considering linear combinations, we simply consider the parities
carried by each qubit as in a complete connectivity case. The trick is
to order the parities in terms of distance to the target qubit
(depending on which qubit they appear) and find a change of basis such
that each canonical vector is at the minimum possible distance from
the target qubit. In other words, the parities at distance $1$ from
the target qubit form a subspace of $\mathbb{F}_2^n$ of rank $k_1$ for
some $k_1$ and we can choose $k_1$ such parities as new canonical
vectors (as long as they are linearly independent). Then the parities
at distance $2$ will add $k_2$ new dimensions, adding $k_2$ new
canonical vectors, etc., until the whole space $\mathbb{F}_2^n$ is
spanned with a new basis. Now, in this new basis, we consider the
components $I \subset \llbracket 1, n \rrbracket$ of the parity to
synthesize that correspond to the farthest canonical vectors. By
construction, such components can only be zeroed by adding parities
from those farthest qubits. Furthermore, adding supplementary parities
on the way will not modify the values the components $I$. So our
proposal therefore is to remove such distant components by greedily
choosing the parities that minimize at each iteration the number of
nonzero components from $I$. To add such parity we choose to use the
template in Fig.~\ref{multi_CNOT_LNN} with the full linear combination
as we said that the intermediate parities do not modify the components
$I$. Once the components $I$ are zeroed, we repeat the process with
the new farthest components until the whole parity is synthesized.

\subsection{Synthesis of a general operator}

The extension of the synthesis from triangular to general operator is
not as straightforward as in the all-to-all connectivity case. We
cannot simply write $A = PLU$ and concatenate the circuits
synthesizing $L$ and $U$ and ultimately permuting the qubits. If we
want to use this algorithm as a sub-task of a global circuit optimizer
for NISQ architectures we cannot afford to swap the qubits because it
could break the optimizations done in the rest of the circuit.

To avoid the permutation of the qubits we have to transform the matrix
$A$ by applying a pre-circuit $C$ such that $CA = LU$.  Then the
concatenation of $C^{-1}$ and the circuits synthesizing $L$ and $U$
gives a valid implementation of $A$.

\subsubsection{Computation of \texorpdfstring{$C$}{C}} 

If $A$ is invertible, which is always the case, then it admits an LU
factorization if and only if all its leading principal minors are
nonzero. We propose an algorithm for computing $C$ exploiting this
property while trying to optimize the final size of $C$. We
successively transform $A$ such that every submatrix $A$[1:i, 1:i] is
invertible. By construction when trying to make $A[1:k,1:k]$
invertible for some $k$ we have $A[$1:$k-1$, 1:$k-1]$ invertible. If
$A[1:k,1:k]$ is invertible then we do nothing, otherwise we look in
the parities $A$[$k+1$:$n$,1:$k$] those who, added to $A[k,1:k]$, make
$A[1:k,1:k]$ invertible. By assumption $A$ is invertible so there is
at least one such row that verifies this property. Then among the
valid parities we choose the closest one to qubit $k$ in $G$. We can
add all the parities along the path because by assumption they belong
to the span of the first $k-1$ rows of $A[1:k,1:k]$ so it has no
effect on the rank of $A[1:k,1:k]$.

\subsubsection{Choice of the qubit ordering} \label{sec::ordering}

We can further optimize our algorithm by changing the qubits
ordering. The algorithm we have presented for synthesizing a
triangular operator is still valid up to row and column
permutations. Thus, given a permutation $P$ of the qubits, one can
synthesize $P^{-1}LP$ by applying our algorithm with the order given
by $P$. Then, instead of computing a circuit $C$ such that $CA = LU$
we search for a circuit $C$ satisfying $P^{-1} CA P = LU$ and
\[ CA = PLP^{-1}PUP^{-1} = L'U' \] where $L'$ and $U'$ can be
synthesized using our algorithm. Searching for such $C$ can be done
using our algorithm on $A[P,P]$ (in Matlab notation, i.e., the
reordering of $A$ along the vector $P$).

This means that we can choose $P$ such that the synthesis of $L$ and
$U$ will yield shorter circuits. Empirically we noticed that when
synthesizing the $k^{\text{th}}$ parity of $L$ it is preferable to
have access to the parities appearing on qubits $k-1, k-2$, etc., in
priority for two reasons: first because they can modify more bits on
the $k^{\text{th}}$ parity and secondly because it is likely that
there will be much more parities available, increasing the chance to
have an inexpensive solution to the weighted syndrome decoding
problem. Intuitively we want the ordering of the qubits to follow at
least an Hamiltonian path in $G = (V,E)$ which would match the
previous restriction on the ordering we formulated at the beginning of
the section.

In practice, we found empirically orderings that provide good results
and in fact they are simple. For instance, for any architecture
similar to a grid, the ordering we chose is the one starting from the
top left qubit, follows the first line, then continues below on the
second line, etc., giving a "snake" structure to the ordering in the
architecture. See Fig.~\ref{architectures} for an illustration on
several architectures.

Finding a good qubit ordering does not have, strictly speaking, to be
automatized because it can be done once for all for a given
architecture. Still, it would be interesting to be able to find good
algorithms for this problem, especially in order to deal with more
complex architectures than the ones we consider in this paper. Some
proposals and discussions are given in Section~\ref{sec::discussion}
to tackle this problem.

\subsubsection{Exploit the symmetries of the architectures} 
Lastly, it is possible to use the symmetries of the architectures to
obtain other qubit orderings that are as valid as the one initially
chosen. For instance, for a grid, starting from one of the three other
corners or following a path along the columns instead of the rows
gives other qubit orderings that can potentially lead to a better
result. We cannot know in advance which ordering will be the most
suited for one particular operator, so the only solution is to try all
possibilities (8 in the case of a grid) and keep the best result.

\subsection{Benchmarks}

\subsubsection{Fine tuning of the method}

By the heuristic nature of our algorithm, its global performance
depends on several parameters:
\begin{itemize}
\item the maximum number of shortest paths between two qubits, noted
  $\text{SP}_{\max}$,
\item the number of linear combination we authorize along a certain
  path, noted $\text{LC}_{\max}$,
\item the method to solve the syndrome decoding problem,
\item the number of times we solve the syndrome decoding problem (with
  random changes of basis), noted $\text{Niter}_{\text{syndrome}}$,
\item the number of times we repeat the synthesis of one triangular
  operator, noted $\text{Niter}$,
\item the ordering of the qubits.
\end{itemize}

The role of each individual parameter is difficult to quantify, but it
is even harder to quantify the interdependent roles of all parameters
together.  After some numerical experiments, we have observed that the
results have a high variance. Notably, the correlation between the
quality of the solutions of the syndrome decoding problem and the
quality of the overall synthesis algorithm is much weaker than in the
all-to-all connectivity case.

It seems that the mean performance of the algorithm cannot be changed
significantly but we can act on the value of the variance of the
results: the larger the variance is, the more likely we will find very
good solutions (among very bad ones). To increase the variance, we
insist on the parameters that add randomness in the results.

Overall, here are our current conclusions on the role of each parameter: 

\paragraph{$\text{Niter}$} This parameter is probably the most
important one. Being able to repeat the experiment a large number of
times increases our chance to find a good solution.

\paragraph{$\text{Niter}_{\text{syndrome}}$} This parameter improves
the solution of the syndrome decoding problem, but has little effect
on the global quality of the results. In fact, it tends to deteriorate
the global performance of our algorithm. This might be due to the fact
that there is less variance in the results, thus less chances to reach
a good solution.

\paragraph{The method to solve the syndrome decoding} Using an integer
programming solver offers almost no variance in the results. We
noticed that the best results were obtained with the greedy
heuristic. When the size of the problem is large, it is preferable to
use the faster heuristic notably due to the computational time.

\paragraph{$\text{LC}_{\max}$} This parameter has little effect on the
results. Furthermore the computational complexity becomes quickly
intractable if we increase $\text{LC}_{\max}$: allowing to remove one
parity from the linear combination given by Eq.~\eqref{eq::lc} adds
$\binom{k}{1} = k$ parities where $k$ is the length of the path
considered, removing two parities adds $\binom{k}{2} \approx k^2/2$
parities, etc. Moreover, those additional parities have a larger cost
and are unlikely to be used in the solution of the syndrome decoding
problem.

\paragraph{$\text{SP}_{\max}$} This parameter has a significant effect
on the results. Contrary to the parameter $\text{LC}_{\max}$, this one
has to be maximized if possible. It introduces more parities (but this
time with identical cost), hence more variance in the results.

\paragraph{The ordering of the qubits} The ordering has a critical
impact on the performance of the algorithm. With some experiments we
come up with some good orderings that give in average the best
results. Exploiting the symmetries also has a non negligible impact as
we increase the chances to find better solutions. More details will be
given in the discussions in Section~\ref{sec::discussion}.

\subsubsection{Numerical results} 

We compare our method against the best algorithm in the literature
\cite{kissinger2020cnot} whose source code is available on the PyZX
Github repository \cite{PyZX}. For each architecture considered in
their implementation we generate a set of 50 random operators and
perform the synthesis using the Steiner trees. Their algorithm
provides an optimization using genetic algorithms but this implements
the circuit up to a permutation of the qubits. We first consider exact
synthesis so in this case we considered their algorithm without this
extra optimization. Then we propose the same experiment but where the
synthesis can be done up to a final permutation of the qubits, in this
case we added the genetic algorithm optimization in the Steiner tree
based algorithm.

Our own algorithm is implemented in Julia. We set a time limit of 10
minutes for the synthesis of an operator. The values of the parameters
for each architecture are summarized in Table~\ref{parameters_values}.
Overall, we solely focused on the number of iterations, the number of
paths and the method to solve the syndrome decoding problem.  When the
number of qubits is small enough ($n \leq 25$) we can maximize the
number of paths considered and the number of iterations with the
standard greedy method.  For large problem sizes ($n \geq 64$), only
the fast heuristic can provide results in a reasonable amount of time
and we modulate the computational time and the performance of our
algorithm with the number of iterations. For intermediate sizes,
namely $n=36$ and $n=49$, the heuristic nature of our algorithm makes
it nontrivial to determine the best set of parameters. From our
observations, we noticed it was best to have a large number of
iterations for the square architectures. However, when adding diagonal
connections between the qubits, having a good tradeoff between the
number of paths and the number of iterations gives better results.

\begin{table}[t]
\caption{Values of the parameters of the syndrome algorithm tuned for different architectures.}
\centering
{\renewcommand{\arraystretch}{1.2}
\scalebox{0.75}{
\begin{tabular}{c|c|cccccc}
\toprule
Architecture & \# & $\text{Niter}$ & $\text{Niter}_{\text{syndrome}}$ & Syndrome solver & $\text{LC}_{\max}$ & $\text{SP}_{\max}$ & Qubit ordering \\
\midrule
9q square  & 9  & 100  & 1 & Greedy  & 1 & Inf & See Fig~\ref{architectures} \\
Rigetti 16q & 16 & 100  & 1 & Greedy  & 1 & Inf & See Fig~\ref{architectures} \\
IBM QX 5 & 16 & 100  & 1 & Greedy  & 1 & Inf & See Fig~\ref{architectures} \\
16q Square & 16 & 100  & 1 & Greedy  & 1 & Inf & See Fig~\ref{architectures} \\
19q Line & 19 & 100  & 1 & Greedy  & 1 & Inf & Natural ordering \\
IBM Q20 Tokyo & 20 & 100  & 1 & Greedy  & 1 & Inf & See Fig~\ref{architectures} \\
25q Square & 25 & 100  & 1 & Greedy  & 1 & Inf & See Fig~\ref{architectures} \\
25q Sq. + diag. & 25 & 100  & 1 & Greedy  & 1 & Inf & See Fig~\ref{architectures} \\
36q Square & 36 & 100  & 1 & Greedy  & 1 & 1 & See Fig~\ref{architectures} \\
36q Sq. + diag. & 36 & 50  & 1 & Greedy  & 1 & 10 & See Fig~\ref{architectures} \\
49q Square & 49 & 100  & 1 & Fast  & 1 & 1 & See Fig~\ref{architectures} \\
49q Sq. + diag. & 49 & 10  & 1 & Greedy  & 1 & 10 & See Fig~\ref{architectures} \\
64q Square & 64 & 50  & 1 & Fast  & 1 & 1 & See Fig~\ref{architectures} \\
81q Square & 81 & 25  & 1 & Fast  & 1 & 1 & See Fig~\ref{architectures} \\
\bottomrule
\end{tabular}}
}
\label{parameters_values}
\end{table}

\begin{table}[t]
\caption{Exact synthesis: performance of our Syndrome Decoding based algorithm vs Steiner trees algorithm \cite{kissinger2020cnot} for several architectures}
\centerline{\scalebox{0.75}{
{\renewcommand{\arraystretch}{1.3}
\begin{tabular}{c|c|c|c|c|c|c|c|c|c} \toprule
  \multirow{2}{*}{Architecture}
  & \multirow{2}{*}{\#}
  & Steiner
  & \multirow{2}{*}{Syndrome}
  & \multicolumn{4}{c|}{Saving}
  & \multirow{2}{*}{$\text{t}_{\text{St}}$ (s)}
  & \multirow{2}{*}{$\text{t}_{\text{Sy}}$ (s)} \\
 &  & \cite{kissinger2020cnot} & & Mean & Min. & Max. &  Positive &  &  \\ \midrule \midrule

9q Square & 9 & 61 & 46 & 23.2\% & 2.33\% & 37.14\% & 100\% & 0.013 & 0.785\\
Rigetti 16q & 16 & 271 & 245 & 9.4\% & 0.8\% & 18.84\% & 100\% & 0.045 & 6.879\\
IBM QX 5 & 16 & 245 & 181 & 25.6\% & 13.08\% & 37.35\% & 100\% & 0.027 & 13.96\\
16q Square & 16 & 206 & 155 & 24.8\% & 10.53\% & 35.78\% & 100\% & 0.019 & 14.106\\
19q Line & 19 & 453 & 454 & -0.3\% & -6.96\% & 10.29\% & 40\% & 0.035 & 9.185\\
IBM Q20 Tokyo & 20 & 294 & 211 & 28.1\% & 22.02\% & 35.14\% & 100\% & 0.025 & 10.385\\
25q Square & 25 & 516 & 397 & 22.9\% & 17.01\% & 27.61\% & 100\% & 0.064 & 213.24\\
25q Square + diag. & 25 & 411 & 299 & 27.3\% & 20.16\% & 32.95\% & 100\% & 0.042 & 78.147\\
36q Square & 36 & 1066 & 865 & 18.8\% & 15.46\% & 23.18\% & 100\% & 0.101 & 277.886\\
36q Square + diag. & 36 & 861 & 645 & 25.0\% & 20.8\% & 29.1\% & 100\% & 0.162 & 328.48\\
49q Square & 49 & 1978 & 1633 & 17.5\% & 13.76\% & 20.21\% & 100\% & 0.407 & 275.378\\
49q Square + diag. & 49 & 1605 & 1230 & 23.3\% & 21.36\% & 26.23\% & 100\% & 0.358 & 350.625\\
64q Square & 64 & 3368 & 2771 & 17.7\% & 15.22\% & 19.69\% & 100\% & 1.043 & 414.579\\
81q Square & 81 & 5372 & 4398 & 18.1\% & 16.7\% & 20.09\% & 100\% & 2.06 & 577.091\\
\bottomrule
\end{tabular}}}}
\vspace{0.5cm}
\label{table_results_exact}
\end{table}

\paragraph{Exact synthesis}

Results for exact synthesis are summarized in
Table~\ref{table_results_exact}. Columns 3 and 4 give the average size
of the generated circuits for the method using Steiner trees in
\cite{kissinger2020cnot} and our algorithm based on syndrome
decoding. The next columns detail the savings: the mean saving, the
minimum saving (negative saving means that our algorithm performs
worse), the maximum saving and the proportion of operators for which
our circuit is actually shorter than the one provided by the
state-of-the-art method. The last two columns give the average time
required to perform the synthesis of one operator (all iterations
included for our algorithm).

We can expect our algorithm to behave better if there are more
connections between the qubits. When the connectivity is as limited as
possible, for instance with an LNN architecture, our algorithm does
not outperform the algorithm based on Steiner trees. The average
performance of both algorithms are almost identical and the natural
distribution of the relative savings are probably only due to the
variance of the results. Similarly, as the Rigetti architecture is
close to a straight line, we outperform the state of the art but with
less savings compared to the other architectures: 10\% of savings in
average but with almost no savings in some cases.

For the remaining architectures the results are more promising. In the
case of the 9-qubit square there is a lot of variance in the results:
depending on the operator we can have a gain of almost 40\% or almost
no savings at all. Overall we still manage to produce a shorter
circuit for every circuit with an average gain of 23\%.

For larger architectures, we outperform the state-of-the-art algorithm
consistently with at least 17\% of savings. First, the more connected
the architecture, the better the results. This is particularly visible
if we add diagonal connections in the square architectures: both
algorithms provide much shorter circuits but we manage to take more
advantage of it, improving our relative savings. We also have almost
30\% savings in average on the IBM-Tokyo chip.

Despite the use of the simpler heuristic for large problems, we still
manage to get at least 17\% of savings in average. This is less than
for smaller architectures for which more optimal techniques are used
but the method is more scalable.

\paragraph{Synthesis up to a permutation}

\begin{table}[t]
  \caption{Synthesis up to a permutation: performance of our Syndrome Decoding based algorithm vs Steiner trees algorithm \cite{kissinger2020cnot} for several architectures}
\centerline{\scalebox{0.75}{
{\renewcommand{\arraystretch}{1.3}
\begin{tabular}{c|c|c|c|c|c|c|c|c|c} \toprule
  \multirow{2}{*}{Architecture}
  & \multirow{2}{*}{\#}
  & Genetic Steiner
  & \multirow{2}{*}{Syndrome}
  & \multicolumn{4}{c|}{Saving}
  & \multirow{2}{*}{$\text{t}_{\text{St}}$ (s)}
  & \multirow{2}{*}{$\text{t}_{\text{Sy}}$ (s)}\\
 &  & \cite{kissinger2020cnot} & & Mean & Min. & Max. &  Positive &  &  \\ \midrule \midrule

9q Square & 9 & 42 & 42 & 0.2\% & -24.32\% & 16.33\% & 54\% & 1.033 & 0.709\\
Rigetti 16q & 16 & 231 & 232 & -0.4\% & -11.37\% & 11.21\% & 42\% & 6.794 & 5.955\\
IBM QX 5 & 16 & 194 & 169 & 12.5\% & 0.0\% & 23.41\% & 98\% & 12.786 & 13.814\\
16q Square & 16 & 167 & 144 & 13.3\% & 4.52\% & 23.12\% & 100\% & 14.961 & 15.662\\
19q Line & 19 & 393 & 434 & -10.7\% & -18.25\% & -1.72\% & 0\% & 15.795 & 9.114\\
IBM Q20 Tokyo & 20 & 250 & 199 & 20.3\% & 12.07\% & 25.77\% & 100\% & 12.61 & 12.046\\
25q Square & 25 & 453 & 381 & 16.0\% & 12.16\% & 20.69\% & 100\% & 244.247 & 211.785\\
25q Square + diag. & 25 & 358 & 284 & 20.5\% & 14.75\% & 25.96\% & 100\% & 67.086 & 75.212\\
36q Square & 36 & 983 & 839 & 14.6\% & 12.15\% & 16.9\% & 100\% & 497.191 & 277.083\\
36q Square + diag. & 36 & 789 & 623 & 21.0\% & 18.37\% & 23.18\% & 100\% & 349.002 & 256.034\\
49q Square & 49 & 1878 & 1597 & 15.0\% & 13.51\% & 16.91\% & 100\% & 356.147 & 275.2\\
49q Square + diag. & 49 & 1512 & 1199 & 20.7\% & 18.33\% & 22.68\% & 100\% & 422.83 & 292.891\\
64q Square & 64 & 3243 & 2725 & 16.0\% & 14.79\% & 17.25\% & 100\% & 560.246 & 434.613\\
81q Square & 81 & 5223 & 4334 & 17.0\% & 15.94\% & 17.79\% & 100\% & 588.478 & 555.365\\

\bottomrule
\end{tabular}}}}
\vspace{0.5cm}
\label{table_results_perm}
\end{table}

In \cite{kissinger2020cnot} the authors propose to synthesize the
operator up to row and column permutations in order to reduce the
total number of CNOT gates and they use a genetic algorithm to find
suitable permutations. Although allowing to permute the rows and the
columns of the operator indeed helps finding shorter circuits, it
cannot be used directly in a peep-hole optimization process for global
circuit optimization.

Still, there are cases where permuting the rows and columns may be
allowed. Given an operator $A$, permuting the columns of $A$ is
equivalent to applying a permutation matrix on the right: this
corresponds in a change of the initial layout of the logical qubits in
the hardware. Modifying the layout of the logical qubits can only be
done once, at the beginning of the optimization. Hence such operation
is not available for all linear reversible operators in the global
circuit. Permuting the rows of $A$ is equivalent to applying a
permutation matrix on its left, this means that the qubits carry the
good parities but are permuted in the hardware. In a context where we
want to optimize on the fly a bigger quantum circuit, this will have
an impact on the remaining circuit to optimize but the optimization
process continues. Therefore permuting the rows might be preferable
over permuting the columns as it can be integrated more easily in a
global optimization algorithm.

We recall that our syndrome decoding based algorithm for restricted
connectivities consists in two parts:
\begin{itemize}
\item First, we compute a short circuit $C$ such that $CA = LU$,
\item Then we synthesize $L, U$ and overall $A = C^{-1}LU$.
\end{itemize}

We know that we can always write $A = PLU$ for some permutation matrix
$P$, so if we authorize ourselves to permute the rows of $A$ we can
simply replace $C$ by a permutation matrix and save the cost of
implementing $C$.

We repeated the experiments done for an exact synthesis but with the
rows permutations. The results are in
Table~\ref{table_results_perm}. We kept the default values of the
parameters of the genetic algorithms given in PyZX. We only modified
the number of iterations to ensure that the computational times of our
method and the Steiner tree based method are approximately the
same. Some finer tuning on the genetic algorithm parameters can be
done and this is a limitation in our comparison. Note also that their
code optimizes both rows and columns permutations, while our algorithm
only allows rows permutations. This is another limitation for a fair
comparison.

Overall, we still manage to outperform \cite{kissinger2020cnot} for
all but three architectures but the savings are not as good as in the
exact synthesis case. Notably for sparse or small architectures our
results are worse (for the 19 qubits line) or equal (for Rigetti's
chip or the 9 qubits square). For the other architectures, the savings
range from 12\% for the IBM QX5 chip to 20\% for the most connected
architectures. Note that the gap we noticed between the small and
large architectures, which was due to the change of the heuristic for
solving the syndrome decoding problem, disappear in this new
experiment. We think this is due to the quality of the genetic
algorithm that struggles optimizing efficiently the rows and columns
permutation when the problem size is too large. This compensates the
loss of quality of our own heuristic.

\paragraph{Experiments with more connected architectures} The
architectures currently available have a sparse qubit connectivity,
making it costly to implement CNOT circuits. Even though full qubit
connectivity will not be technologically feasible anytime soon, we can
hope that more connected architectures will be designed in a near
future. Will this favor our algorithm? In our experiments, we
highlighted the fact that the more connected the architecture, the
better the savings over the Steiner tree based algorithm. Besides, the
closer we get to a full connectivity the more our algorithm will
perform similarly to our original syndrome decoding based algorithm
while the Steiner tree based will eventually consist in a standard
Gaussian elimination. This is another argument in favor of an
increasing outperformance of our algorithm over the state of the
art. We propose two experiments to confirm or infirm this behavior.

The first experiment is based on an experimental model that can
already be found in certain technologies such as some using Rydberg
atoms \cite{pasqal}. Starting from a grid layout, we consider that
each qubit has a fixed interaction radius and can interact with the
qubits within its reach. Here the radius is computed with the
$\text{L}_2$ norm in a plane. So for instance setting the radius to
$1$ and we have the standard square architecture. Setting the radius
to $\sqrt{2}$ and we recover the square architecture with diagonal
connections.

\begin{figure}
\centering
\includegraphics[scale=0.55]{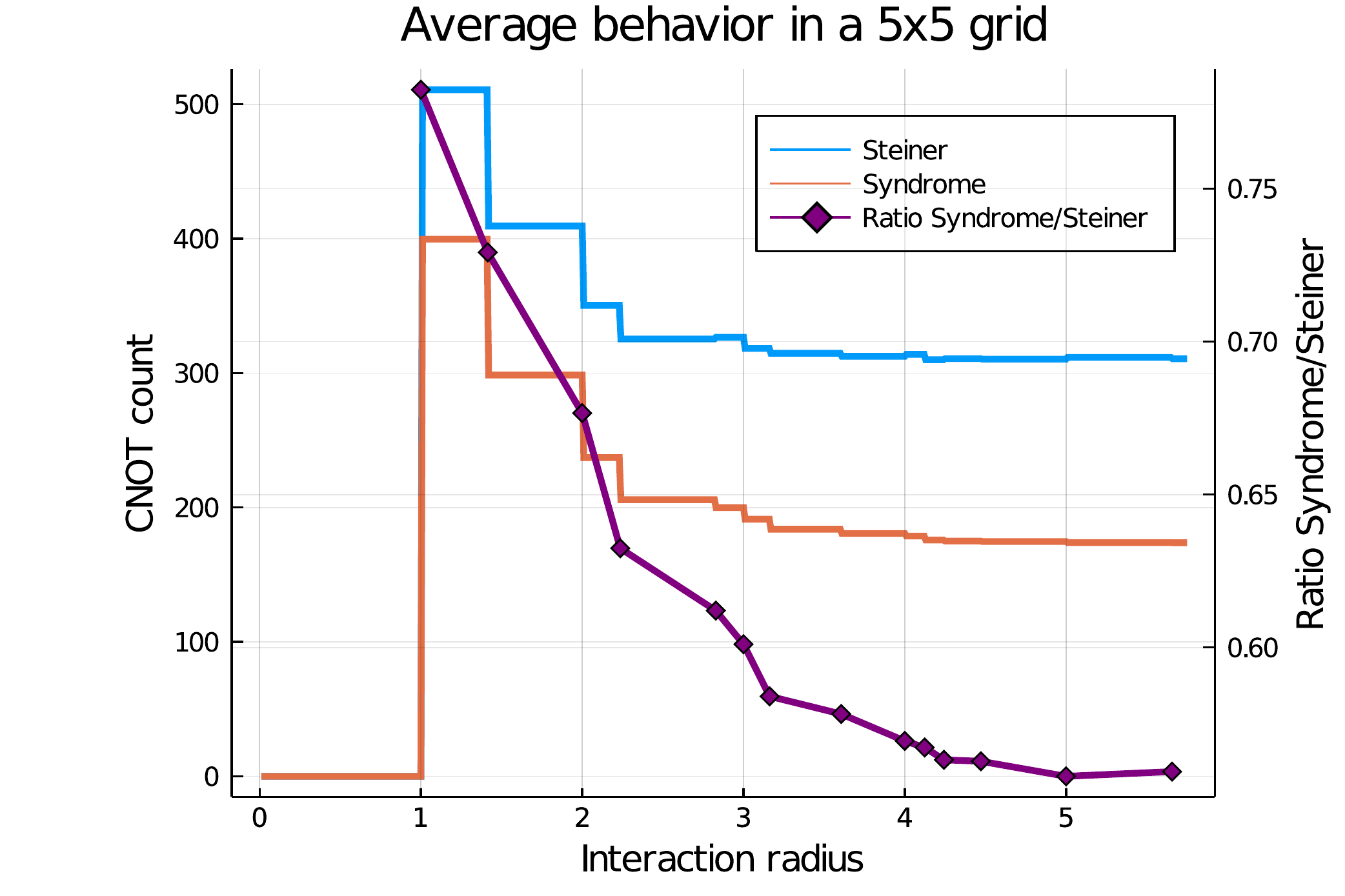}
\caption{Average CNOT count for the synthesis of CNOT circuits in a 5x5 grid with increasing interaction radius.}
\label{grid_exp}
\end{figure}

For a $25$-qubit square, one can show that the number of interactions
increases for the following values of the radius:
\[ 1, \sqrt{2}, 2, \sqrt{5}, \sqrt{8}, 3, \sqrt{10}, \sqrt{13}, 4,
  \sqrt{17}, \sqrt{18}, \sqrt{20}, 5, \sqrt{32} \] and for each of
these radius we computed the average CNOT count of both our algorithm
and the state of the art with a sample of $50$ operators. The results
are given in Fig.~\ref{grid_exp}. The behavior of the two algorithms
is similar: the CNOT count decreases exponentially with the
interaction radius. This is very promising because it shows that only
a small improvement in the interaction radius can lead to consequent
savings in the cost of CNOT circuits. Interestingly, the difference
between the CNOT counts of the two algorithms is approximately
constant. Necessarily this results in an increasing relative gain of
our algorithm against the state-of-the-art method.

\begin{figure}
\includegraphics[scale=0.55]{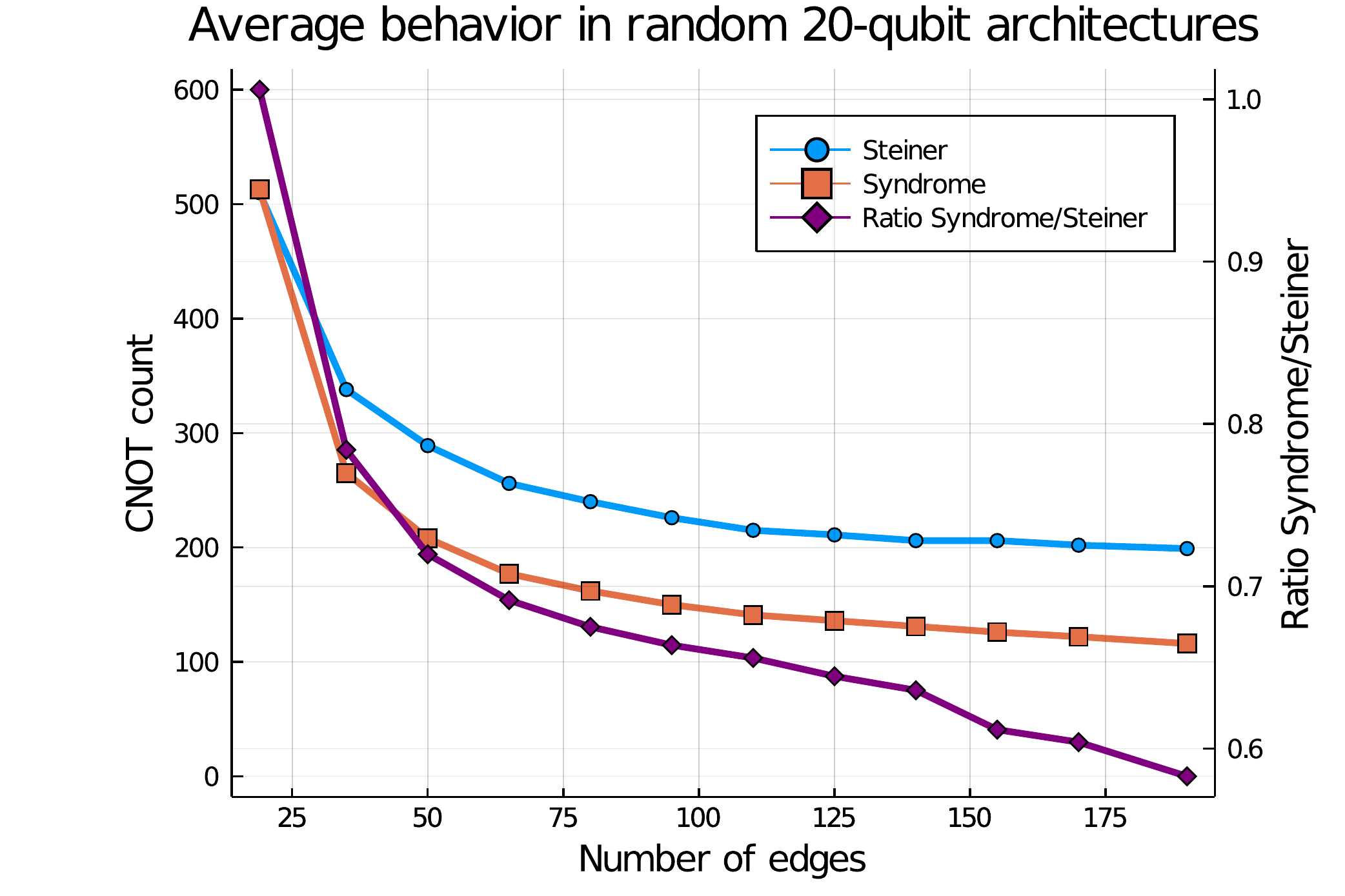}
\caption{Average CNOT count for the synthesis of CNOT circuits in increasingly connected random 20-qubit architectures.}
\label{LNN_exp}
\end{figure}

The second experiment aims at showing that the advantage of our
algorithm in more connected architectures is still true even in an
unstructured architecture. Starting from a 20-qubit LNN architecture,
we iteratively randomly add $15$ edges to our architecture for a total
of $12$ increasingly connected architectures. For each new
architecture we synthesize $50$ random operators and we store the
average CNOT count. The results are given in Fig~\ref{LNN_exp}. We get
similar results to the experiment with the grid and the increasing
interaction radius.

\subsection{Discussion} \label{sec::discussion}

We now discuss a crucial problem for the performance of the syndrome
decoding algorithm: the choice of the qubit ordering. Indeed, we
noticed empirically that a good qubit ordering has a non negligible
impact on the overall performance of our algorithm. Unfortunately, the
role of the qubit ordering is quite opaque. We propose intuitive
mathematical formulations of the best qubit ordering and we discuss
their validity.

We define an ordering as a map
$\pi : V \to \llbracket 1,n \rrbracket$, and the distance between two
qubits $u,v \in V$ in the ordering is given by $| \pi(u)-\pi(v)|$. We
also note $d(u,v)$ the distance between qubits $u$ and $v$ in the
hardware.

In Section~\ref{sec::ordering}, we defined a good qubit ordering as an
order of the qubits such that each qubit is the closest possible in
the hardware to its predecessors and successors in the ordering. In
other words, we want at least two successive qubits in the ordering to
be neighbors in the hardware. Qubits at distance $2$ in the ordering
should also be at distance $2$ or less in the hardware, etc. A
Hamiltonian path in the hardware guarantees that
$d(u,v) \leq \|\pi(u)-\pi(v)\|$, but among the Hamiltonian paths in
the hardware, some may be better than others.

\paragraph{The minLA problem} In \cite{RC20}, we initially formulated
the best ordering $\pi : V \to \llbracket 1,n \rrbracket$ as a
solution of the Minimum Linear Arrangement problem
\begin{equation} 
\begin{aligned} 
  & \underset{\pi}{\text{minimize}}
  & \sum_{(u,v) \in E} w_{uv} \; |\pi(u)-\pi(v)| \\
\end{aligned}
\end{equation}
where $w_{uv}$ is the weight of the edge connecting $u$ and $v$ in the
graph. The idea is that we want to give priority to neighbors in the
hardware: the nodes must be as close as possible in the hardware if
their ``numbers'' are also close. A way to do so is to solve the MinLA
problem, not in the hardware graph, but in the complete graph with
suitable weights. Namely $w_{ij}$ must be large when $i,j$ are
neighbors in the hardware and $w_{ij}$ must be smaller if $i,j$ are at
distance 2, and even smaller for larger distances.
The MinLA problem has already been used for qubit routing
\cite{pedram2016layout} and the problem is in general
NP-Hard~\cite{garey1979computers}. One way to solve exactly this
problem is to encode it in an integer linear program. For conciseness,
we will not detail the formulation of the problem but is uses standard
encoding techniques. The important result is that, with a time limit
given to our integer programming solver, we are able to find solutions
quickly, although those solutions are not proven to be optimal.

Unfortunately, the solutions are in fact quite deceiving, some of them
not even being a Hamiltonian path of the hardware graph. We believe
this is not because the solver did not find optimal solutions, but
rather because the minLA problem is in fact not the most suitable
problem to encode what we want. The reason is the following: the
minimum value of $|\pi(u) - \pi(v)| = 1$ can only be assigned to $n-1$
possible combinations of the $(u,v)$s. On the other hand, large
weights are given to every pair of neighbor nodes in the hardware and
there are as many as there are edges in the hardware graph. Even for
sparse architectures, like a square, there are approximately four
times more maximum weights $w_{ij}$ than minimum possible assignments
of $|\pi(u) - \pi(v)|$. In other words, most of the cost to minimize
will consist in the portion
\begin{equation}
  \text{cste} \times \sum_{u,v \text{ neighbors}}
  |\pi(u)-\pi(v)|. \label{eq::order}
\end{equation}

Given that we sum over a large number of terms, some of the
$|\pi(u)-\pi(v)|$ will not be equal to $1$ but rather $2, 3$ or even
more. In these circumstances, as long as the total sum is minimized,
the optimizer will not necessarily find a solution where
$|\pi(u)-\pi(v)| = 1$ will necessarily correspond to neighbor qubits
$(u,v)$. We illustrate this with the example given in
Table~\ref{ordering_minLA}. We give two different orderings for which
Eq.~\eqref{eq::order} is minimized but one of them do not correspond
to a Hamiltonian path. The two orderings have the same cost for any
weights $(w_{uv})_{uv}$ and one can show that this cost is minimal for
interesting values of the weights (for instance
$w_{uv} = 1/(10)^{d(u,v)}$).

\paragraph{A nonlinear integer programming formulation.} As a response
to the limitation of the minLA problem, we propose a new way to encode
our problem of qubit ordering by changing the cost function. To
emphasize the role of the small $|\pi(i)-\pi(j)|$ terms, we have to
use a nonlinear function. We set
\[ w_{uv} = d(u,v)\] and the function to minimize is now
\begin{equation}
  \sum_{uv} w_{uv} e^{-|\pi(u)-\pi(v)|}.\label{nonlinear}
\end{equation}
With this cost function, we can see that the terms that have to be
minimized in priority are the cases where $|\pi(u)-\pi(v)| = 1$ so
they will have to correspond to qubits $(u,v)$ that correspond to
small $w_{uv}$. In other words, for neighbor qubits $(u,v)$ we will
have $|\pi(u)-\pi(v)| = 1$. And the logic is the same for qubits that
are at distant 2, 3, etc.

\begin{figure}
\centering
\includegraphics[scale=0.5]{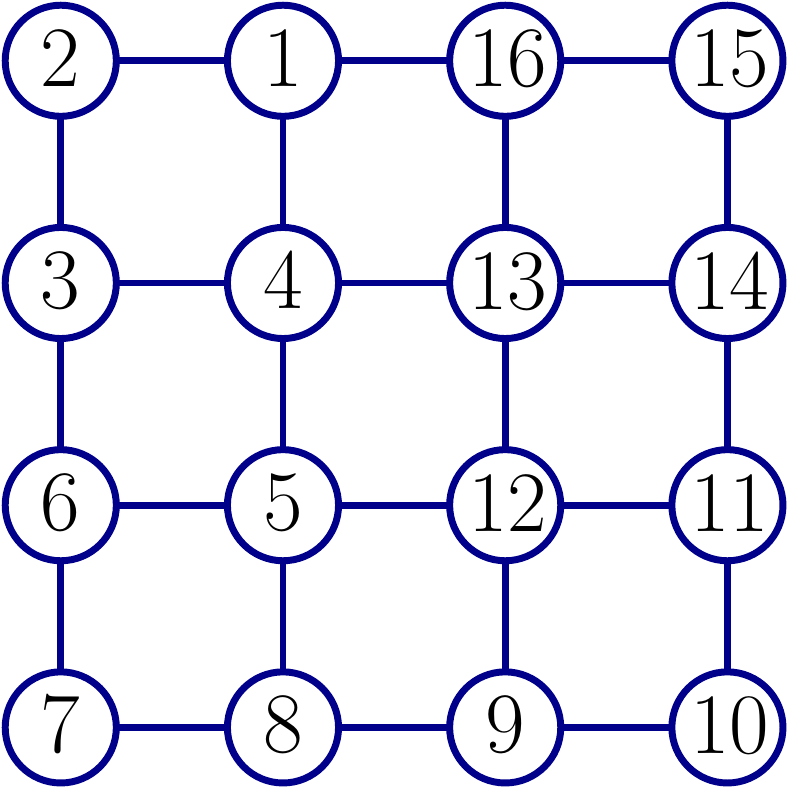}
\caption{Ordering given by our heuristic minimizing Eq.~\eqref{nonlinear}.}
\label{ordering_heuristic}
\end{figure}

This problem can be similarly cast into a nonlinear integer
programming. Unfortunately, the solvers at our disposal struggle
finding solutions. We have to use an heuristic. We propose this very
simple algorithm: we choose a starting permutation and at each
iteration we switch two elements of our permutation. We choose one
pair among the ones that minimize the cost function and we stop when a
local minimum is found. Finally we repeat this process with different
random starting permutations. An example of ordering obtained with
this method for the 16-qubit square is given in
Fig.~\ref{ordering_heuristic}. In theory such ordering seems better
suited than the standard "snake" we used in our benchmarks. If we
check, the cost function is indeed smaller for this ordering than for
the snake. It makes sense because of the winding pattern of the
ordering. However, in practice, this does not improve our results (165
CNOT in average against 155). We do not really have an explanation, it
is probably due to the fact that our initial definition of a good
ordering, i.e., an order of the qubits such that each qubit is the
closest possible in the hardware to its predecessors and successors in
the ordering, does not catch the whole complexity of the problem. Does
there exist a better formulation of the perfect qubit ordering?

\begin{table}
\centerline{\scalebox{0.75}{
{\renewcommand{\arraystretch}{1.5}
\begin{tabular}{c|c|c}
\toprule
\toprule
Solver & Empirical & MinLA + Integer Programming \\
& & \\
Ordering & \begin{minipage}{.8in}\includegraphics[width=.8in]{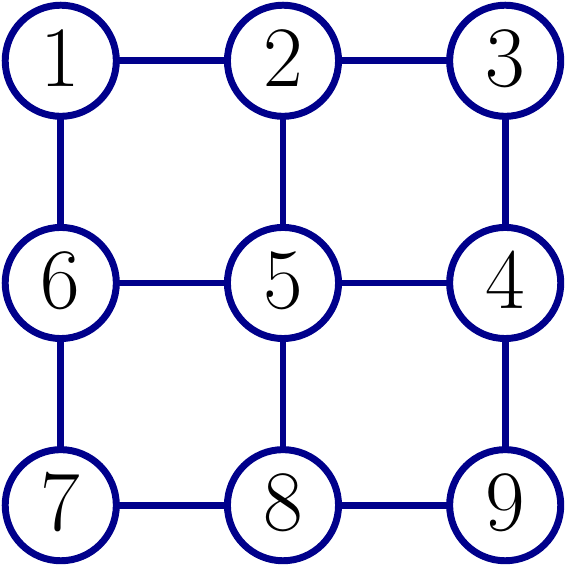}\end{minipage} &  \begin{minipage}{.8in}\includegraphics[width=.8in]{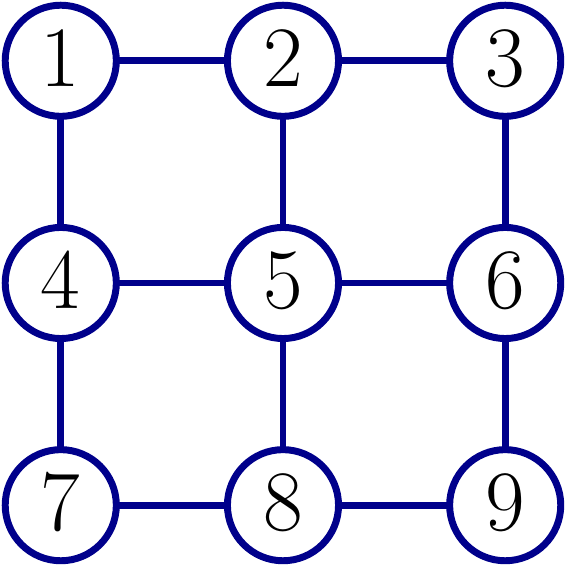}\end{minipage} \\
\midrule
\midrule
\multirow{2}{*}{$\displaystyle \sum_{\substack{u,v \\ d(u,v)=1}} | \pi(u) - \pi(v) |$} & $|1-2| + |1-6| + |2-3| + |2-5| + |3-4| + |4-5|+$  & $|1-2| + |1-4| + |2-3| + |2-5| + |3-6| + |4-5|+$ \\
& \hspace*{0.4cm} $|4-9| + |5-6| + |5-8| + |6-7| + |7-8| + |8-9| = 24$ & \hspace*{0.4cm} $|4-7| + |5-6| + |5-8| + |6-9| + |7-8| + |8-9| = 24 $ \\
\hline
\multirow{3}{*}{$\displaystyle \sum_{\substack{u,v \\ d(u,v)=2}} | \pi(u) - \pi(v) |$} & $|1-3| + |1-5| + |1-7| + |2-4| + |2-6| + |2-8| +$ & $|1-3| + |1-5| + |1-7| + |2-4| + |2-6| + |2-8| +$\\
& \hspace*{-0.1cm} $|3-5| + |3-9| + |4-6| + |4-8| + |5-7| + |5-9| + $ & \hspace*{-0.4cm} $ |3-5| + |3-9| + |4-6| + |4-8| + |5-7| +|5-9|$ \\
& \hspace*{0.4cm} $|6-8| + |7-9| = 48$ & $|6-8| + |7-9| = 48$\\
\hline
\multirow{2}{*}{$\displaystyle \sum_{\substack{u,v \\ d(u,v)=3}} | \pi(u) - \pi(v) |$} & $ |1-4| + |1-8| + |2-7| + |2-9| + |3-6| + |3-8| +$ & $ |1-6| + |1-8| + |2-7| + |2-9| + |3-4| + |3-8| + $\\
 & \hspace*{0.4cm} $|4-7| + |6-9| = 36$ & $ |4-9| + |6-7| = 36$ \\
\hline
$\displaystyle \sum_{\substack{u,v \\ d(u,v)=4}} | \pi(u) - \pi(v) |$ & $|1-9| + |3-7| = 12 $ & $|1-9| + |3-7|  =12 $ \\
\midrule
Total cost & $12w[1] + 48w[2] + 36w[3] + 12w[4]$& $12w[1] + 48w[2] + 36w[3] + 12w[4]$ \\
\bottomrule
\end{tabular}}}}
\caption{Ordering given empirically versus a solution of the minLA problem. Both orderings have the minimal possible cost but one is not a Hamiltonian path.}
\label{ordering_minLA}
\end{table}

\section{Conclusion} \label{conclusion} We have presented a new
framework for the synthesis of linear reversible circuits. We exploit
the specific structure of triangular operators to transform the
synthesis into a series of syndrome decoding problems, which are
well-known problems in cryptography. Using an LU decomposition we can
synthesize any quantum operator in the case of an all-to-all
connectivity. Benchmarks show that we outperform the state-of-the-art
algorithm for intermediate sized problems ($n < 400$). Our heuristics
for solving the syndrome decoding problem are efficient but could
still be improved, both in circuit size and computational time. For
instance, some quantum algorithms have been proposed for solving the
syndrome decoding problem via the Information Set Decoding algorithm
\cite{bernstein2010grover,kachigar2017quantum,kirshanova2018improved},
which gives the possibility of designing a hybrid quantum/classical
compiler for this particular synthesis problem.

Then we have highlighted the robustness of our framework by extending
it to an arbitrary connectivity graph having a Hamiltonian path. With
a suitable pre-processing of the matrix we transform the problem into
a series of weighted syndrome decoding problems. Except for the LNN
architecture whose connectivity is too sparse, we consistently
outperform existing algorithms.  As a future work, we can study how to
extend our method to the case where the connectivity graph does not
have a Hamiltonian path, similarly to~\cite{kissinger2020cnot}. For
the moment we only have studied the behavior of our algorithm on
random CNOT circuits, but large-scale CNOT circuits are not common. It
would be interesting to extend our framework in order to deal with
quantum circuits implementing real algorithms, e.g., quantum
chemistry-based circuits or arithmetic functions. Besides, with
specific quantum circuits instead of random CNOT circuits, the
outperformance of direct synthesis methods is not guaranteed and a
more in-depth study of various methods (direct synthesis, SWAP
insertions, etc.) will be necessary.

\section*{Acknowledgment}
This work was supported in part by the French National Research Agency
(ANR) under the research project SoftQPRO ANR-17-CE25-0009-02, and by
the DGE of the French Ministry of Industry under the research project
PIA-GDN/QuantEx P163746-484124. We thank Bertrand Marchand for
comments on the manuscript.

\end{document}